\documentclass[letterpaper,nomarginnotes,nonarrowgutter]{jpaper}

\usepackage[sort,compress]{cite}

\usepackage{xspace}
\usepackage{array}
\usepackage{adjustbox}
\usepackage{listings}
\usepackage{setspace}

\definecolor{gfored}{rgb}{0.580, 0.050, 0.211}
\definecolor{ao}{rgb}{0.007, 0.520, 0.867}
\definecolor{moegi}{rgb}{0.357, 0.537, 0.188}
\definecolor{jl}{rgb}{1.0, 0.2, 0.8}
\definecolor{brown(web)}{rgb}{0.65, 0.16, 0.16}
\definecolor{bisque}{rgb}{1.0, 0.89, 0.77}
\definecolor{nbs}{rgb}{0.88, 0.07, 0.37}
\definecolor{yt}{rgb}{0.58, 0.44, 0.86}
\definecolor{iy}{rgb}{0.0, 0.56, 0.041}
\definecolor{mel}{rgb}{0.9, 0.55, 0.31}
\definecolor{ouscolor}{rgb}{0.0, 0.2, 0.4}
\definecolor{fe}{rgb}{0.2, 0.85, 0.75}
\definecolor{frenchblue}{rgb}{0.19, 0.55, 0.91}

\newcommand{\exploitingRowHammerAllCitations}[0]{\cite{gruss2016rowhammer, fournaris2017exploiting, poddebniak2018attacking, tatar2018throwhammer, carre2018openssl, barenghi2018software, zhang2018triggering, bhattacharya2018advanced, google-project-zero, kim2014flipping, rowhammergithub, seaborn2015exploiting, van2016drammer, razavi2016flip, pessl2016drama, xiao2016one, bosman2016dedup, bhattacharya2016curious, burleson2016invited, qiao2016new, brasser2017can, jang2017sgx, aga2017good, mutlu2017rowhammer, tatar2018defeating, gruss2018another, lipp2018nethammer, van2018guardion, frigo2018grand, cojocar2019eccploit,  ji2019pinpoint, mutlu2019rowhammer, hong2019terminal, kwong2020rambleed, frigo2020trrespass, cojocar2020rowhammer, weissman2020jackhammer, zhang2020pthammer, yao2020deephammer, deridder2021smash, hassan2021utrr, jattke2022blacksmith, tol2022toward, kogler2022half, orosa2022spyhammer, zhang2022implicit, liu2022generating, cohen2022hammerscope, zheng2022trojvit, fahr2022frodo, tobah2022spechammer, rakin2022deepsteal, kang2024sledgehammer, jattke2024zenhammer, bolcskei2025rubicon, deridder2025posthammer, jattke2025mcsee, meyer2026phoenix, lin2025gpuhammer, lin2026gpubreach, hu2026gddrhammer, wan2026geforge, shukla2026prowhammer, chen2025rhammer, li2025oneflip}}

\usepackage[most]{tcolorbox}
\tcbset{before skip=4.1pt, after skip=4.1pt}

\definecolor{charcolor}{HTML}{4A90A4}   %

\newtcolorbox[auto counter]{tkx}[2][]{%
  enhanced, breakable,
  colframe  = #2!60!black,
  colback   = #2!5,
  boxrule   = 0.6pt,
  arc       = 1.5pt,
  left      = 4pt,
  right     = 4pt,
  top       = 3pt,
  bottom    = 3pt,
  #1%
}

\newcounter{char}
\newcommand\characteristic[1]{%
  \refstepcounter{char}%
  \begin{tkx}{charcolor}%
    \noindent\textbf{\sffamily Characteristic~\thechar.} #1%
  \end{tkx}%
}

\definecolor{mechcolor}{HTML}{2E8B57}     %

\newcounter{mech}
\newcommand\matchingmechanism[1]{%
  \refstepcounter{mech}%
  \begin{tkx}{mechcolor}%
    \noindent\textbf{\sffamily Matching Device-Level Mechanism~\themech.} #1%
  \end{tkx}%
}

\definecolor{unmechcolor}{HTML}{CD5C5C}   %

\newcounter{unmech}
\newcommand\unmatchingmechanism[1]{%
  \refstepcounter{unmech}%
  \begin{tkx}{unmechcolor}%
    \noindent\textbf{\sffamily Unmatching Device-Level Mechanism~\theunmech.} #1%
  \end{tkx}%
}

\definecolor{takecolor}{HTML}{7B68EE}    %

\newcounter{take}
\newcommand\takeaway[1]{%
  \refstepcounter{take}%
  \begin{tkx}{takecolor}%
    \noindent\textbf{\sffamily Takeaway~\thetake.} #1%
  \end{tkx}%
}

\definecolor{obscolor}{HTML}{E07B39}    %
\newcounter{obs}
\newcommand\observation[1]{%
  \refstepcounter{obs}%
  \begin{tkx}{obscolor}%
    \noindent\textbf{\sffamily Observation~\theobs.} #1%
  \end{tkx}%
}

\providecommand{\Description}[1]{}

\usepackage{fancyhdr}
\usepackage[bookmarks=true,breaklinks=true,letterpaper=true,colorlinks,citecolor=blue,linkcolor=blue,urlcolor=blue]{hyperref}

\title{{\huge Demystifying DRAM Read Disturbance:\\Bridging the Gap Between Experimental Characterization and Device-Level Modeling of RowHammer and RowPress Phenomena}}

\renewcommand{\thefootnote}{\fnsymbol{footnote}}

\author{
Haocong Luo$^{1}$\thanks{These authors contributed equally to this work.}\quad
Longda Zhou$^{2}$\footnotemark[1]\quad
Ataberk Olgun$^{1}$\quad
{\.I}smail~Emir~Y{\"u}ksel$^{1}$\\[0.3em]
Nisa Bostanci$^{1}$\quad
Zhigang Ji$^{3}$\quad
Xing Wu$^{2}$\quad
Onur Mutlu$^{1}$\\[0.4em]
{\textit{$^{1}$ETH Zurich\quad
$^{2}$East China Normal University\quad
$^{3}$Shanghai Jiao Tong University}}
}

\begin{document}
\bstctlcite{IEEEexample:BSTcontrol}

\maketitle
\thispagestyle{fancy}

\renewcommand{\thefootnote}{\arabic{footnote}}
\setcounter{footnote}{0}

\begin{abstract}
    DRAM read disturbance, like RowHammer and RowPress, is a critical robustness issue where accessing DRAM can cause unintended bitflips in other unaccessed DRAM locations. DRAM read disturbance bitflips significantly impact the safe, secure, and reliable operation of DRAM-based computing systems. Many prior works extensively perform experimental characterization of DRAM read disturbance bitflips and propose mitigation techniques based on the empirical characterization results. Some other device-level works study the underlying physical mechanisms of DRAM read disturbance bitflips, but these studied mechanisms do not fully explain all the empirical observations from experimental characterization studies.

    Our goal in this paper is to bridge the gap between experimental characterization and device-level modeling and understanding of RowHammer and RowPress to provide a principled foundation for future works on understanding, characterizing, and mitigating DRAM read disturbance. To this end, we first identify and demonstrate the gaps and inconsistencies between the current understanding of the physical mechanisms of RowHammer and RowPress from existing device-level modeling works and experimental characterization of RowHammer and RowPress bitflips. We focus on three most fundamental metrics of RowHammer and RowPress read disturbance bitflips that should map to the first-order underlying physical mechanisms: 1) bitflip directions, 2) the bitflip counts, and 3) the minimum number of aggressor row activations to trigger the first bitflips (i.e., $\mathrm{AC}_{min}$). Second, we present a comprehensive and rigorous set of TCAD simulations that match the observed phenomena from experimental characterizations of RowHammer and RowPress bitflips. 
    
    From our results, we 1) summarize a set of updated device-level error mechanisms for understanding RowHammer and RowPress bitflips, and 2) identify key modeling and simulation parameters that significantly affect whether simulation results match real-chip characterization. We discuss the implications of our findings on 1) rigorous, comprehensive, and efficient experimental characterization methodologies of DRAM read disturbance bitflips, and 2) the designs of DRAM read disturbance mitigation techniques.
\end{abstract}

\section{Introduction}
Read disturbance is a critical robustness issue in DRAM where accessing DRAM rows (i.e., aggressor rows) can cause \emph{unintended} bitflips in nearby \emph{unaccessed} DRAM rows (i.e., victim rows). DRAM read disturbance phenomena like RowHammer~\cite{kim2014flipping,kim2020revisiting} and RowPress~\cite{luo2023rowpress,luo2024experimental} significantly impact the safe, secure, and reliable operation of DRAM-based computing systems. Malicious parties can exploit DRAM read disturbance bitflips to break the fundamental security property of memory isolation~\exploitingRowHammerAllCitations{}.

Many prior works extensively perform experimental characterization on real DRAM chips~\cite{kim2014flipping,kim2020revisiting,orosa2021deeper,yaglikci2022understanding,luo2023rowpress,yaglikci2024svard,nam2024dramscope} to 1) understand read disturbance bitflips under different conditions like data pattern, access pattern, temperature, and 2) measure key robustness metrics like the minimum number of aggressor row activations to induce at least one bitflip ($AC_{min}$) and the bit error rate (BER). DRAM read disturbance mitigation techniques rely on these empirical characterization results and measurements to be properly configured to ensure the robustness of their designs. 

To fundamentally understand DRAM read disturbance phenomena and then eventually fully mitigate DRAM read disturbance bitflips, it is important to understand the underlying device-level error mechanisms of DRAM read disturbance bitflips. Prior works have identified two major device-level mechanisms that cause DRAM read disturbance: 1) parasitic coupling between the aggressor and victim wordlines, 2) electron migration from near the aggressor DRAM cell to the victim DRAM cell in the silicon substrate. However, the device-level error mechanisms derived from these prior works fail to \emph{fully} explain all the major observations and characteristics of DRAM read disturbance bitflips from experimental characterization works. A prior work~\cite{luo2025revisiting} performs a preliminary study on identifying the gaps and inconsistencies between the existing device-level studies and a subset of experimental observations of DRAM read disturbance bitflips. For example,~\cite{luo2025revisiting} finds that although the state-of-the-art device-modeling of double-sided RowHammer only produces 1-to-0 bitflips, real chip characterization results consistently show both 0-to-1 and 1-to-0 bitflips.

Our goal in this paper is to bridge the gap between real chip characterization and device-level modeling of DRAM read disturbance. We focus on RowHammer and RowPress because they are the two most prominent DRAM read disturbance phenomena. First, we summarize the key device-level error mechanisms of RowHammer and RowPress from prior device-level modeling works~\cite{ryu2017overcoming,yang2019trap, walker2021ondramrowhammer, Gautam2020MitigatingPassing, zhou2023double}. Based on these works, we identify how 1) the DRAM cell array layout, and 2) the DRAM cell structure affects key RowHammer and RowPress bitflip characteristics. 

Second, we design real-DRAM chip characterization experiments to reproduce the RowHammer and RowPress bitflip characteristics identified from prior device-level modeling works. By carefully analyzing the characterization results and mapping them to the DRAM array layout and cell structures, we identify 1) which device-level mechanisms studied in prior works are consistent with real-chip characterization results, and 2) the gaps between the remaining device-level mechanisms and real-chip characterization results.

Third, we bridge these identified gaps and inconsistencies with rigorous and comprehensive device-level TCAD simulations that fully explain the experimental observations. We also identify key device-level modeling parameters that are critical to aligning device-level modeling with experimental observations. For example, we find that fully explaining the bitflip directions of double-sided RowHammer requires modeling 1) the different distances of the two aggressor wordlines to the victim DRAM cell and 2) the structural heterogeneity of the charge trap distribution. We also find that the bulk hole-trap density critically determines whether device-level RowPress modeling matches real-chip characterization results. From our modeling and simulation, we summarize an updated and detailed set of underlying physical error mechanisms to explain the RowHammer and RowPress bitflip characteristics. 

Based on our results, we discuss the implications of our findings on 1) rigorous, comprehensive, and efficient experimental characterization methodologies of DRAM read disturbance bitflips, and 2) the designs of DRAM read disturbance mitigation techniques. For example, we find that the bit error rate (BER) at a given (usually high) aggressor row activation count, although easier to characterize, is not a direct proxy of the minimum aggressor activation count to induce at least one bitflip ($\mathrm{AC}_{min}$) for double-sided RowHammer because 0-to-1 and 1-to-0 bitflips have sophisticated underlying error mechanisms that behave differently.

We make the following key contributions in this paper:
\begin{itemize}
    \item We extract and summarize the key DRAM read disturbance bitflip characteristics observed in prior empirical experimental characterization studies.
    \item We compare existing device-level studies to the experimental observations and identify the gaps and inconsistencies between them.
    \item We bridge the gaps and inconsistencies with rigorous and comprehensive device-level TCAD simulations to 1) understand the underlying physical error mechanisms, and 2) identify key device-level modeling parameters that are critical to aligning device-level modeling with experimental observations.
    \item Our work serves as the foundation for future works to understand, characterize, and mitigate DRAM read disturbance in a principled and efficient manner.
\end{itemize}

\section{Background}
\subsection{Basic DRAM Organization and Operation}
\begin{figure}[ht]
    \centering
    \includegraphics[width=0.75\linewidth]{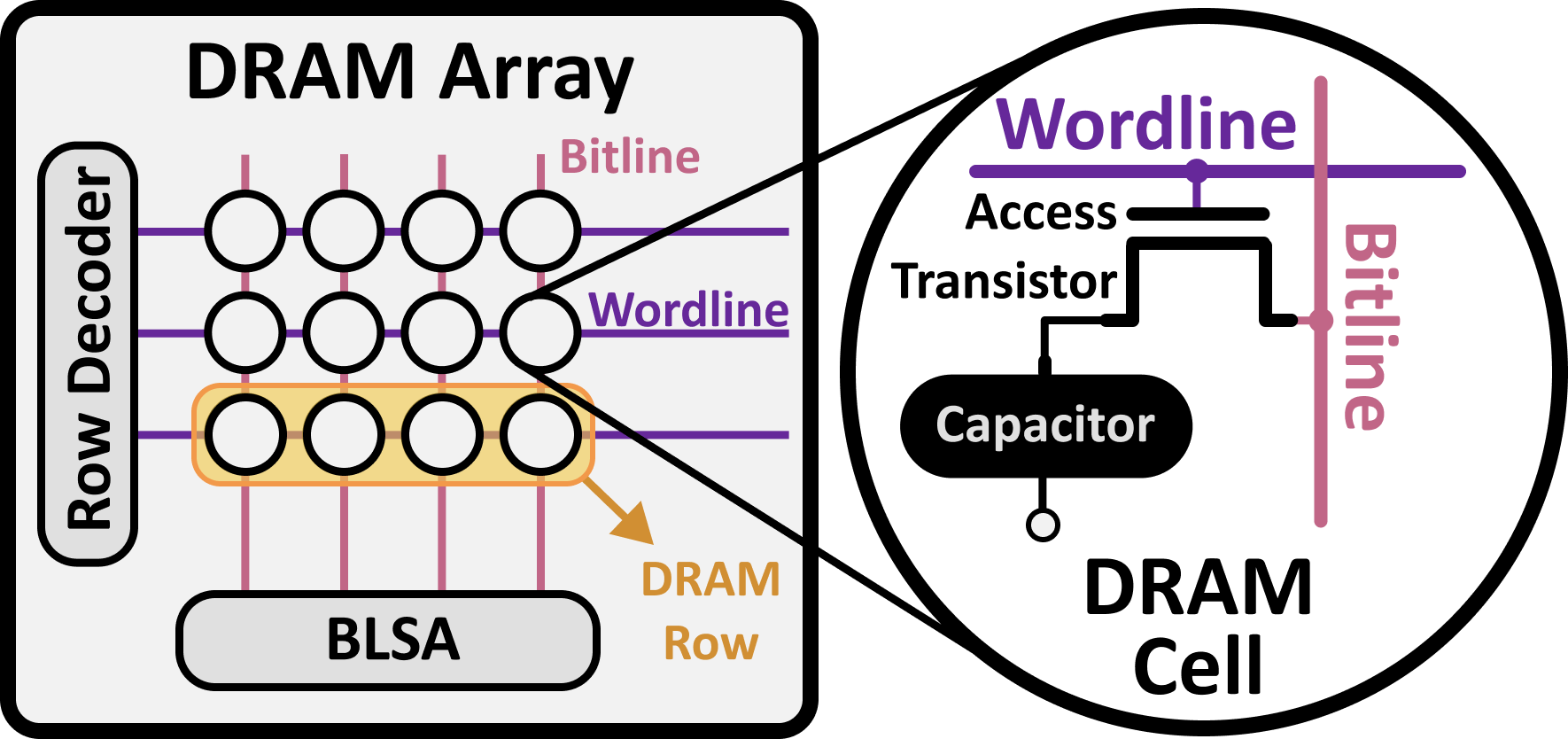}
    \caption{Logical organization of DRAM.}
    \label{fig:dram_org_background}
    \Description{Logical organization of DRAM.}
\end{figure}

Figure~\ref{fig:dram_org_background} shows the \emph{logical organization} of DRAM. DRAM stores data in the form of electric charge in the capacitor of each DRAM cell. DRAM cells are organized into a 2D array, and indexed by rows and columns. A row of DRAM cells share the same wordline (WL), which controls their respective access transistors. 

To access data in DRAM, the memory controller first sends an ACT (activate) command that enables the wordline of a DRAM row. The WL then turns on the access transistors of all DRAM cells in the activated row, connecting the capacitors to the bitlines (BLs). The BLs are connected to the bitline sense amplifiers (BLSA). The BLSA latches the data from the DRAM cells that the memory controller can use read/write (RD/WR) commands to access. After the memory controller finishes accessing the data, it sends a PRE (precharge) command that closes the activated row to prepare the DRAM for the next access. 
\subsection{Physical DRAM Cell Layout and Structure}
\label{sec:dram_layout_background}
\begin{figure}[ht]
    \centering
    \includegraphics[width=\linewidth]{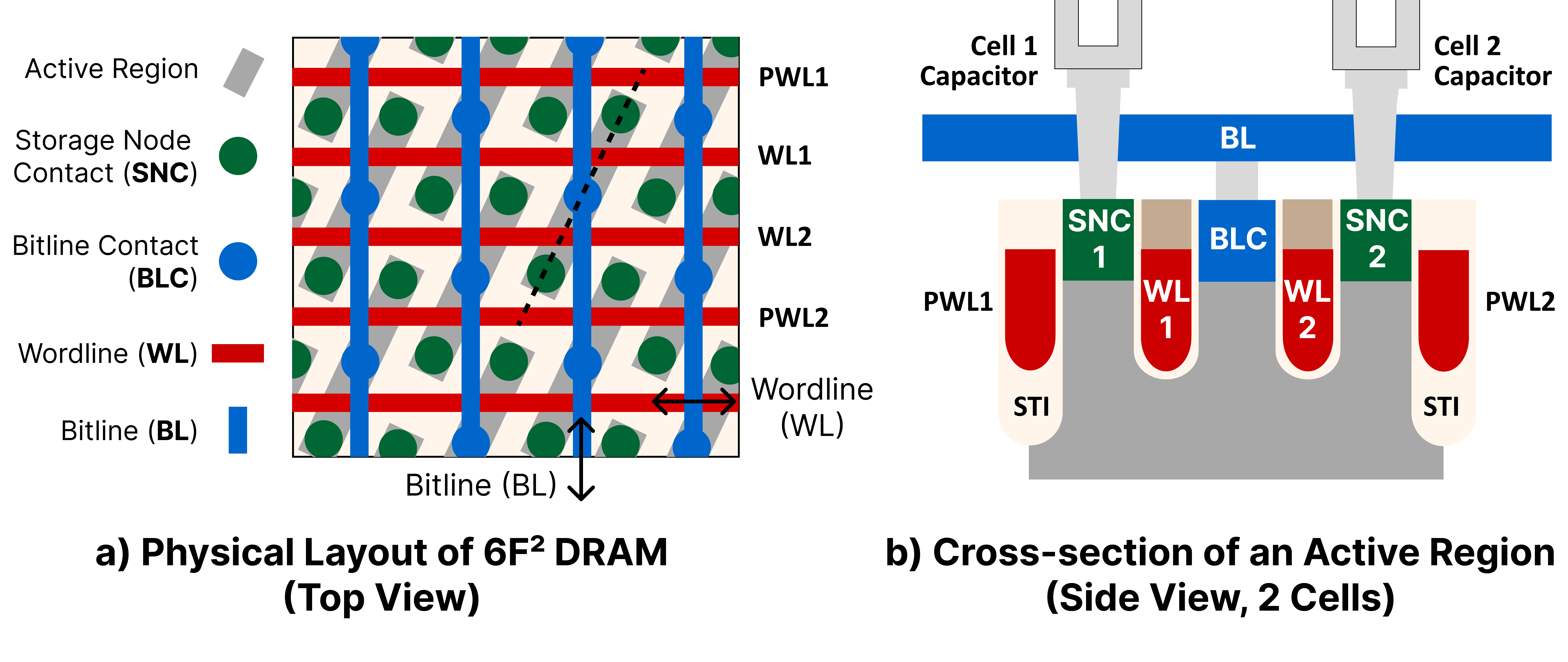}
    \caption{Physical layout of DRAM cells showing a) the top view of a $\mathrm{6F^2}$  DRAM cell array, and b) the cross-sectional side view from the black dashed line in a) showing two adjacent DRAM cells sharing the same active region.}
    \label{fig:dram_layout_background}
    \Description{Physical layout of DRAM cells.}
\end{figure}

Figure~\ref{fig:dram_layout_background}.a) illustrates the top view of the \emph{physical layout} of DRAM cells in a modern high density $\mathrm{6F^2}$ array. In a $\mathrm{6F^2}$ DRAM cell array, two DRAM cells in two adjacent rows that share the same BL also share the same active region (i.e., the physical region where the access transistors are formed). 

Figure~\ref{fig:dram_layout_background}.b) illustrates the cross-sectional side view of such two DRAM cells sharing the same active region from the black dashed line in a). We denote the storage node contacts (SNC, i.e., where the cell capacitor connects to the access transistor) of the two DRAM cells as SNC1 and SNC2. The two cells share the same bitline contact (BLC, i.e., their respective access transistors also share one terminal). For such two DRAM cells, four wordlines are of significance: 1) the two wordlines of the two DRAM cells themselves (i.e., WL1 and WL2), and 2) the two wordlines that do not go over but ``pass by'' the active region (i.e., passing wordlines PWL1 and PWL2). Without loss of generality, for the rest of the paper, we denote cell 1 as the victim cell, cell 2 as the aggressor cell, WL2 as the neighboring aggressor wordline (NWL), and PWL as the passing aggressor wordline (PWL).

Due to the alternating cell-wordline layout in the $\mathrm{6F^2}$ array layout, the same aggressor row will be the NWL for half of the cells in the victim row, and also be the PWL for the other half of the cells in the same victim row. For example, without loss of generality, if the upper aggressor row is the NWL for all the even-indexed cells in the victim row, then the same upper aggressor row will also be the PWL for all the odd-indexed victim cells. The lower aggressor row will then be the PWL for all the even-indexed victim cells and the NWL for all the odd-indexed cells.

\subsection{DRAM Read Disturbance}
We study two major DRAM read disturbance phenomena identified and extensively characterized in prior works: RowHammer~\cite{kim2014flipping,kim2020revisiting} and RowPress~\cite{luo2023rowpress,luo2024experimental}.

\noindent\textbf{RowHammer.} Repeatedly activating and closing the aggressor DRAM row(s) \emph{many times} (e.g., tens of thousands of times) disturbs and causes bitflips in adjacent victim rows.

\noindent\textbf{RowPress.} Keeping the aggressor row(s) activated for a \emph{long period of time} disturbs and causes bitflips in adjacent victim rows. RowPress {requires} much fewer aggressor row activations to cause bitflips compared to RowHammer.

\section{Existing Device-Level Modeling of RowHammer and RowPress}
\label{sec:existing_device_level}
In this section, we summarize the key device-level mechanisms of RowHammer and RowPress extracted from prior device-level modeling works~\cite{ryu2017overcoming,yang2019trap, walker2021ondramrowhammer, Gautam2020MitigatingPassing, zhou2023double, Zhou2024Unveiling, Zhou2024Understanding}.
\subsection{Single-Sided RowHammer}
\begin{figure}[ht]
    \centering
    \includegraphics[width=\linewidth]{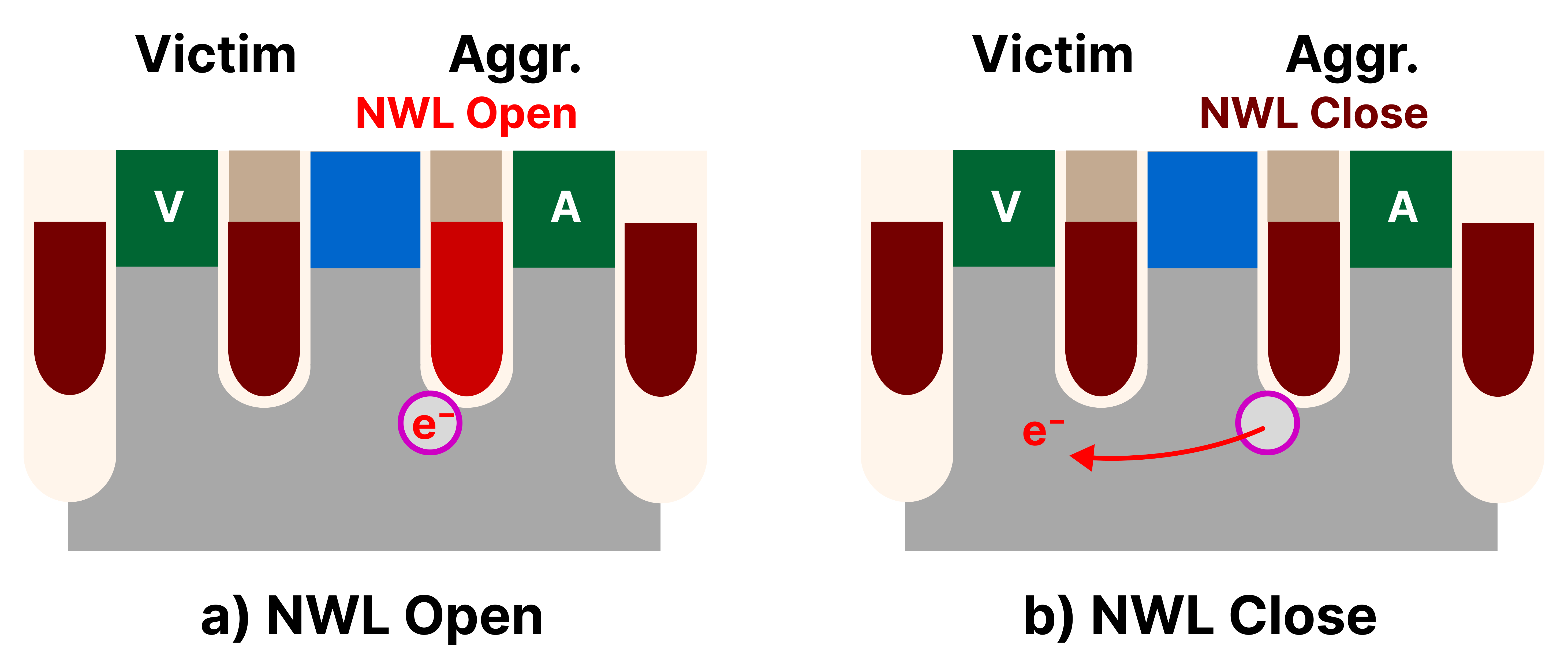}
    \caption{Device-Level mechanism of NWL single-sided RowHammer inducing 1-0 bitflips.}
    \label{fig:nwl-rh}
    \Description{Device-Level mechanism of NWL single-sided RowHammer inducing 1-0 bitflips.}
\end{figure}

For single-sided RowHammer, the aggressor wordline can be either the neighboring wordline (NWL) or the passing wordline (PWL). Figure~\ref{fig:nwl-rh} illustrates the high-level device-level mechanism of NWL-induced RowHammer as shown by prior works~\cite{ryu2017overcoming,yang2019trap, walker2021ondramrowhammer, zhou2023double}. When the NWL is open, charge traps near the NWL capture electrons (Figure~\ref{fig:nwl-rh}.a). When the NWL is closed, the captured electrons are emitted and some of them migrate to the victim storage node (Figure~\ref{fig:nwl-rh}.b), causing its potential to decrease and eventually induce 1-to-0 bitflips.

\characteristic{Repeated opening and closing of the NWL injects electrons to the victim cell, causing 1-to-0 bitflips.}

\begin{figure}[ht]
    \centering
    \includegraphics[width=\linewidth]{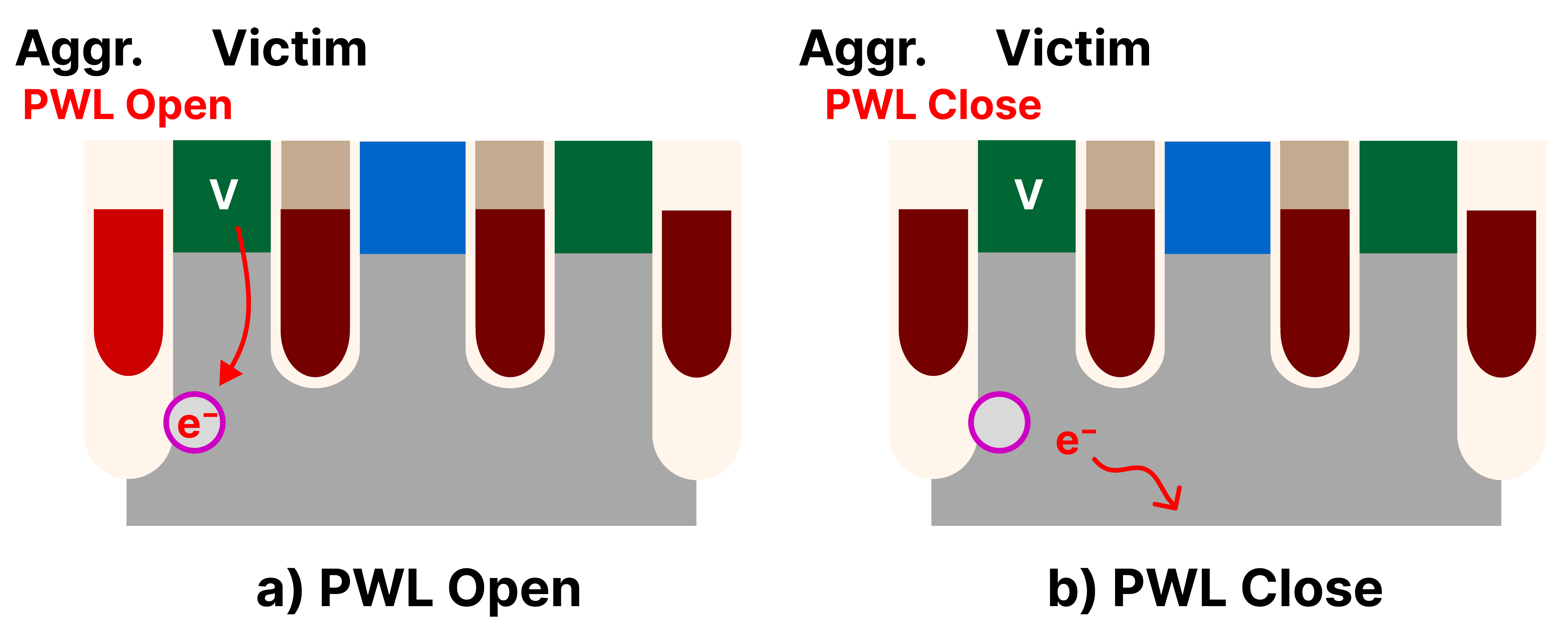}
    \caption{Device-Level mechanism of PWL single-sided RowHammer inducing 0-1 bitflips.}
    \label{fig:pwl-rh}
    \Description{Device-Level mechanism of PWL single-sided RowHammer inducing 0-1 bitflips.}
\end{figure}

Figure~\ref{fig:pwl-rh} illustrates the high-level device-level mechanism of PWL-induced RowHammer as shown by prior works~\cite{Gautam2020MitigatingPassing, zhou2023double}. When the PWL is open, its electric field pulls electrons away from the victim storage node (Figure~\ref{fig:pwl-rh}.a). When the PWL is closed, not all the pulled-away electrons return to the storage node (Figure~\ref{fig:pwl-rh}.b, some of the electrons go towards the P-well of the transistor, some others go to the bitline contact), causing the storage node potential to increase and eventually induce 0-to-1 bitflips.

\characteristic{Repeated opening and closing of the PWL pulls electrons away from the victim cell, causing 0-to-1 bitflips.}
\subsection{Double-Sided RowHammer}
\begin{figure}[ht]
    \centering
    \includegraphics[width=\linewidth]{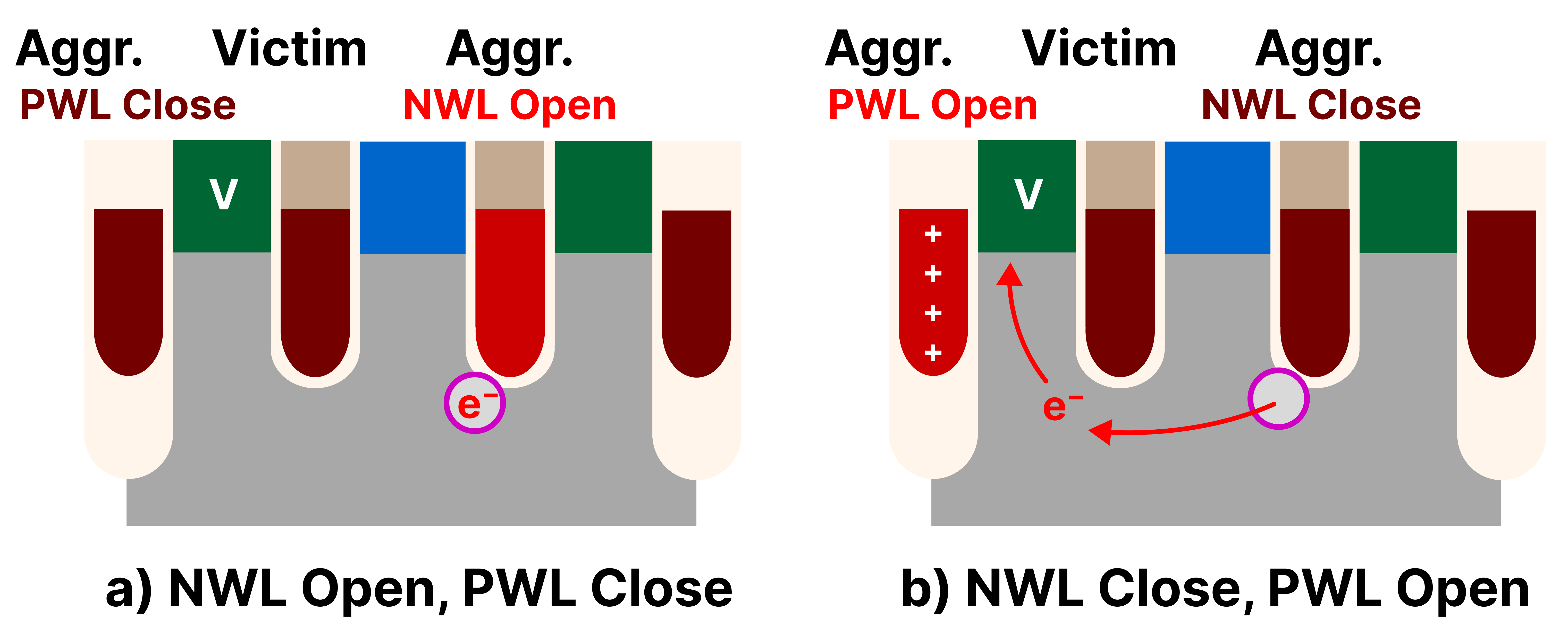}
    \caption{Device-Level mechanism of double-sided RowHammer enhancing 1-to-0 bitflips.}
    \label{fig:ds-rh}
    \Description{Device-Level mechanism of double-sided RowHammer enhancing 1-to-0 bitflips.}
\end{figure}

Double-sided RowHammer involves alternating opening and closing of both the NWL and the PWL. Without loss of generality, we assume NWL is opened first. Figure~\ref{fig:ds-rh} illustrates the device-level mechanism of Double-Sided RowHammer that explains why double-sided RowHammer needs much fewer aggressor row activations than single-sided RowHammer to induce the first bitflip from the state-of-the-art device-level study~\cite{zhou2023double}. When the NWL is open (i.e., the PWL is closed), the charge traps near the NWL capture electrons (Figure~\ref{fig:ds-rh}.a). When the NWL is closed (i.e., the PWL is open), the emitted electrons are pulled towards the victim direction by the strong electric field caused by the opened PWL. This electric field enhances the 1-to-0 bitflips by injecting more electrons into the victim storage node. The simulation results in~\cite{zhou2023double} show that double-sided RowHammer suppresses 0-to-1 bitflips because the kinetics of trapped charge near the PWL is insensitive to the opening and closing of the NWL. 
\characteristic{For double-sided RowHammer, PWL opening during NWL close enhances the electron injection into the victim cell, enhancing 1-to-0 bitflips and suppressing 0-to-1 bitflips.}

\subsection{RowPress}
\begin{figure}[ht]
    \centering
    \includegraphics[width=\linewidth]{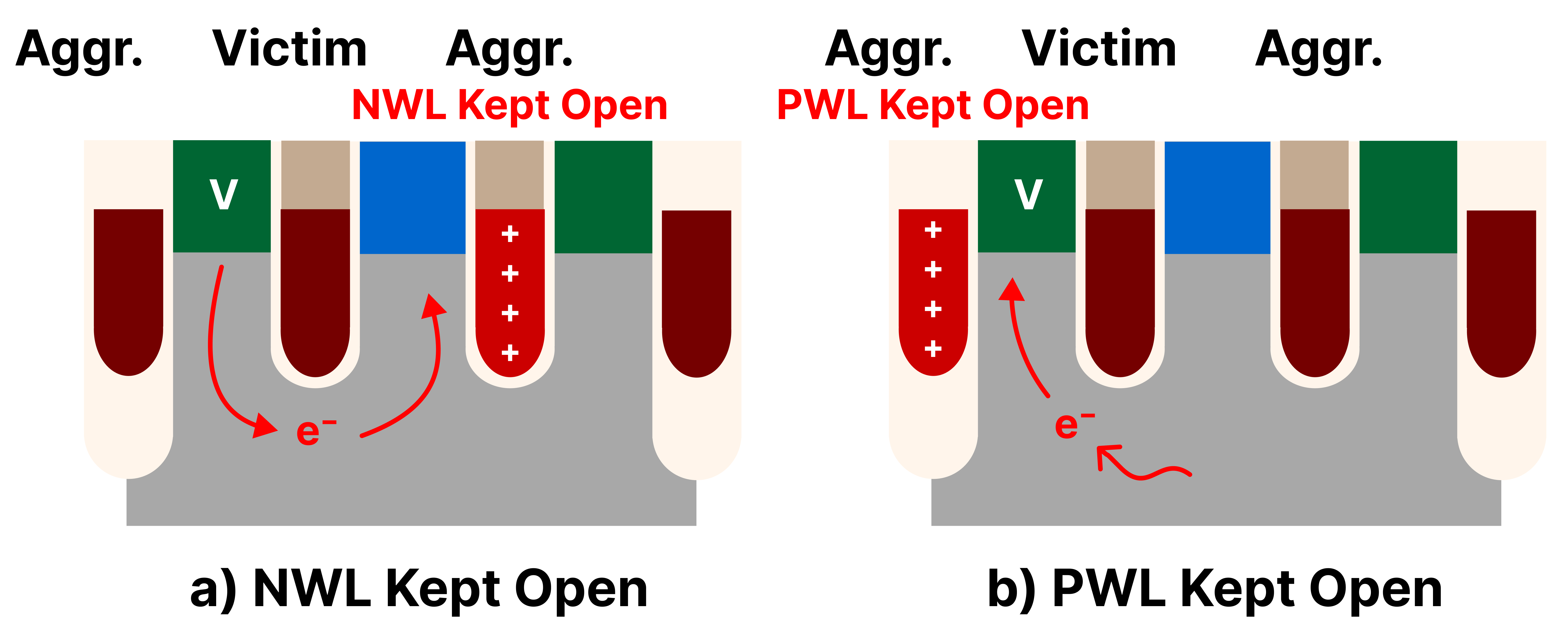}
    \caption{Device-Level mechanism of a) NWL-induced RowPress and b) PWL-induced RowPress.}
    \label{fig:rp}
    \Description{Device-Level mechanism of a) NWL-induced RowPress and b) PWL-induced RowPress.}

\end{figure}

Figure~\ref{fig:rp} illustrates the high-level device-level mechanism of a) NWL-induced RowPress, and b) PWL-induced RowPress from state-of-the-art device-level modeling of RowPress~\cite{Zhou2024Understanding, Zhou2024Unveiling}. 

When the NWL is kept open for a long period of time, its strong electric field 1) weakens the gate control of the victim cell's access transistor, and 2) draws
electrons away from the victim storage node towards the NWL. Both enhance electron migration from the victim to the bitline contact, causing 0-to-1 bitflips. When the PWL is kept open for a long period of time, its strong electric field draws electrons from the transistor substrate towards the PWL direction and some of these electrons are injected into the victim storage node, causing 1-to-0 bitflips.

Compared to RowHammer, for the same aggressor WL type, RowPress induces \emph{the opposite} bitflip directions. This is because 1) the charge traps saturate quickly after the aggressor WL is opened, and 2) the actual RowHammer induced leakage happens when the aggressor WL is closed. As the aggressor WL keeps open (i.e., RowPress), the RowPress induced leakage accumulates over time and eventually overshadows the RowHammer induced leakage at the end.

\characteristic{As tAggON increases (i.e., from RowHammer to RowPress), the bitflip direction induced by NWL (PWL) changes from 1-to-0 (0-to-1) to 0-to-1 (1-to-0).}

\section{Key Characteristics of DRAM Read Disturbance Bitflips from Experimental Characterization}
\label{sec:exp_char}
\subsection{Methodology}
We perform experimental characterization using real commodity off-the-shelf DDR4 DRAM chips to understand the key characteristics of RowHammer and RowPress bitflips using DRAM Bender~\cite{olgun2023dram} with fine-grained control of DRAM command, timing, and temperature. Table~\ref{tab:DRAMs} lists the DRAM modules we test. For each module, we randomly sample and test 128 victim rows. We reverse-engineer 1) the DRAM internal row mapping, and 2) the true- and anti-cell layout of all the tested DRAM modules and account for the layout when initializing the aggressor and victim rows.

\begin{table}[htbp]
\caption{DRAM Chips Tested}
\label{tab:DRAMs}

\centering

\resizebox{\columnwidth}{!}{%
\begin{tabular}{@{}c|c|cc|c|c|c@{}}
\toprule
{\textbf{Mfr.}} & 
{\textbf{Module Type}} & 
{\textbf{Die Density}} & 
{\textbf{Die Revision}} & 
{\textbf{DQ}} & 
{\textbf{Num. Chips}}& 

\begin{tabular}[c]{@{}c@{}}\textbf{Date Code}\\\textbf{(YYWW)}\end{tabular}
 \\ \midrule
S & UDIMM & 8 Gb  & D & $\times$8 & 8 & 2110 \\
S & UDIMM & 16 Gb & M & $\times$8 & 8 & 2118 \\
S & UDIMM & 16 Gb & A & $\times$8 & 8 & 2319 \\
S & UDIMM & 16 Gb & B & $\times$8 & 8 & 2315 \\
S & UDIMM & 16 Gb & C & $\times$8 & 8 & 2408 \\ \midrule
H & UDIMM & 8 Gb  & C & $\times$8 & 8 & 2120 \\
H & UDIMM & 8 Gb  & D & $\times$8 & 8 & 1938 \\
H & UDIMM & 16 Gb & A & $\times$8 & 8 & 2003 \\
H & UDIMM & 16 Gb & C & $\times$8 & 8 & 2136 \\ \midrule
M & UDIMM & 8 Gb  & E & $\times$8 & 8 & 2402 \\ \bottomrule
\end{tabular}%
}

\end{table}

\subsection{Single-Sided RowHammer}
\label{sec:ss-rh-characteristics}
First, we characterize 1) the bitflip directions, 2) the spatial distribution of the bitflips, and 3) the relationship among the bitflip direction, spatial distribution, and the aggressor row location of single-sided RowHammer. We initialize the victim row with either all 0s or all 1s, and always initialize both the upper and lower aggressor rows\footnote{We define the upper and lower aggressor rows as the two rows with physical offset (i.e., after DRAM internal row mapping) +1 and -1, with respect to the victim row, respectively.} to have the opposite data pattern as the victim row. We hammer the aggressor row as many times as possible within the refresh windows of 64ms to 1) maximize the observed bitflips, and 2) avoid observing any retention failure bitflips.

\begin{figure}[ht]
    \centering
    \includegraphics[width=\linewidth]{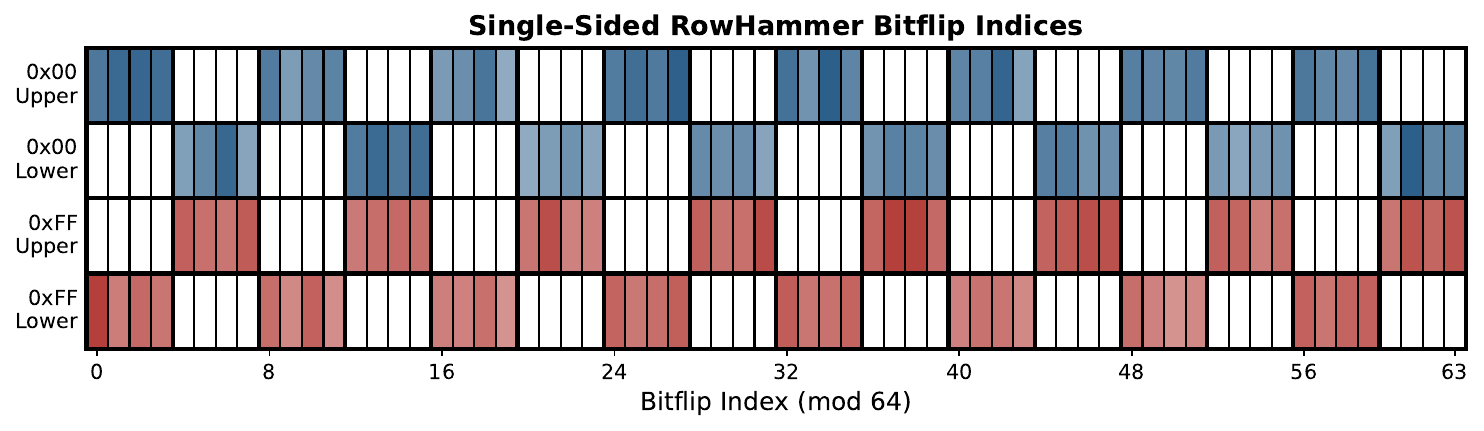}
    \caption{Bitflip direction and spatial distribution of single-sided RowHammer from an example module.}
    \label{fig:rh-bfdir}
    \Description{Bitflip direction and spatial distribution of single-sided RowHammer from an example module.}
\end{figure}

Figure~\ref{fig:rh-bfdir} shows the spatial distribution heatmap of the bitflip indices of single-sided RowHammer from an example DRAM module we test. Each blue (red) cell represents the existence of 0-to-1 (1-to-0) bitflips at a bit location in the victim row. We observe a consistent and regular bitflip spatial pattern that repeats for every 64 bits: 1) the 0-to-1 bitflips always appear in \emph{disjoint} and \emph{alternating} positions compared to the 1-to-0 bitflips, and 2) for the same bitflip location, if it shows 0-to-1 bitflips with the upper aggressor row, then it always shows 1-to-0 bitflips with the lower aggressor row (vice versa).

For all other modules we test, although the exact patterns may differ, we can always find a consistent, regular, and repeating (for every 64 or 512 bits) spatial bitflip pattern for all victim rows we test that matches the following two observations.

\observation{For the same victim bitflip direction, the upper and lower aggressor rows always induce bitflips in alternating and disjoint bit positions.}
\observation{For the same aggressor row location, the 0-to-1 and 1-to-0 bitflips are always in alternating and disjoint bit positions.}

We find that these spatial patterns are consistent with 1) the physical layout of the $\mathrm{6F^2}$ DRAM cell array, and 2) the device-level mechanisms of single-sided RowHammer summarized in Section~\ref{sec:existing_device_level}. First, for the same aggressor wordline, the $\mathrm{6F^2}$ DRAM cell array layout guarantees that for every other cell in the victim row, it is either the NWL or the PWL. For example, consider a victim row and its upper aggressor row. If the upper aggressor row is the PWL for all even-indexed victim cells, then the same aggressor row is the NWL of all the odd-indexed victim cells. Second, for single-sided RowHammer, the two aggressor row locations (i.e., NWL or PWL) induce \emph{opposite} bitflip directions. 

Using these observations, for all modules we test and for every single bit in the victim row, we reverse engineer whether the upper (lower) aggressor row is the victim bit's NWL or PWL. We crosscheck the reverse-engineered NWL and PWL mapping with the bitflip directions and find that almost all DRAM modules we test show 100\% matching results (the only exception is the 16Gb C-Die module from Mfr. S with 99.7\% matching, which we attribute to remapped columns from repair).

From these observations, we conclude that the following two device-level mechanisms match experimental characterization.

\matchingmechanism{Single-Sided NWL RowHammer induces 1-to-0 bitflips.}
\matchingmechanism{Single-Sided PWL RowHammer induces 0-to-1 bitflips.}

\subsection{Double-Sided RowHammer}
We extract and summarize three key characteristics of double-sided RowHammer from the comprehensive and rigorous experimental characterization results of a prior work~\cite{luo2025revisiting}.

\observation{Double-sided RowHammer induces both 0-to-1 and 1-to-0 bitflips.}
\observation{The $\mathrm{AC}_{min}$ of 0-to-1 bitflips is significantly smaller than that of 1-to-0 bitflips in double-sided RowHammer.}
\observation{As both aggressors are hammered for a sufficiently high amount of times, more DRAM cells are vulnerable to 1-to-0 bitflips than 0-to-1 bitflips.}

As~\cite{luo2025revisiting} shows, existing device-level mechanisms studied in state-of-the-art device-level modeling of double-sided RowHammer are not consistent with our observations. First, existing device-level mechanisms show suppressed 0-to-1 bitflips for double-sided RowHammer while experimental characterization shows 1) double-sided RowHammer induces both 0-to-1 and 1-to-0 bitflips, and 2) the error mechanism for double-sided RowHammer induced 0-to-1 bitflips is initially stronger than that of 1-to-0 bitflips. Existing device-level mechanisms also do not explain why we observe more 1-to-0 bitflips than 0-to-1 bitflips only when both aggressor rows are hammered for a sufficiently high amount of times.

\unmatchingmechanism{Prior device-level modeling shows double-sided RowHammer enhances 1-to-0 bitflips but suppresses 0-to-1 bitflips. However, experimental characterization shows double-sided RowHammer 1) induces both 0-to-1 and 1-to-0 bitflips, and 2) the error mechanism for 0-to-1 bitflips is initially stronger than that of 1-to-0 bitflips.}

\subsection{RowPress}
\begin{figure}[ht]
    \centering
    \includegraphics[width=\linewidth]{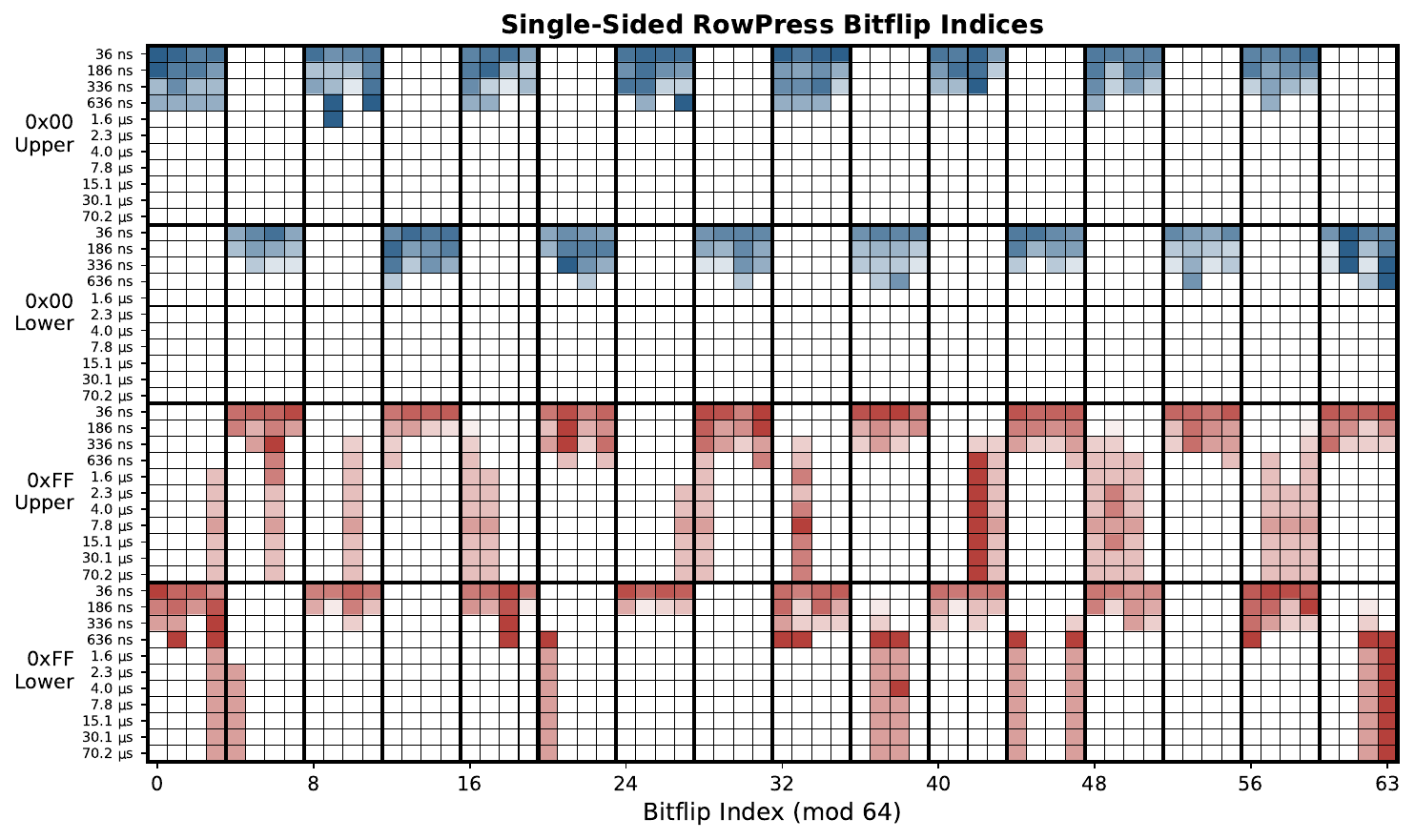}
    \caption{Bitflip direction and spatial distribution of single-sided RowPress.}
    \label{fig:rp-bfdir}
    \Description{Bitflip direction and spatial distribution of single-sided RowPress.}
\end{figure}

Figure~\ref{fig:rp-bfdir} shows the spatial distribution heatmap of single-sided RowPress bitflips from the same example DRAM module as in Section~\ref{sec:ss-rh-characteristics}. We test the DRAM chips at 80$^\circ$C because prior works show that DRAM is more vulnerable to RowPress at a higher temperature~\cite{luo2023rowpress}. The y-axis represents different aggressor row on time (tAggON, increasing order from top to bottom), the x-axis is the bitflip index. We make the following observations from the figure.
\observation{RowPress does not induce 0-to-1 bitflips under normal DRAM operating conditions.}
\observation{For the same aggressor row location, RowPress mainly induces 1-to-0 bitflips in DRAM cells that do not show RowHammer 1-to-0 bitflips.}

\begin{figure}[ht]
    \centering
    \includegraphics[width=\linewidth]{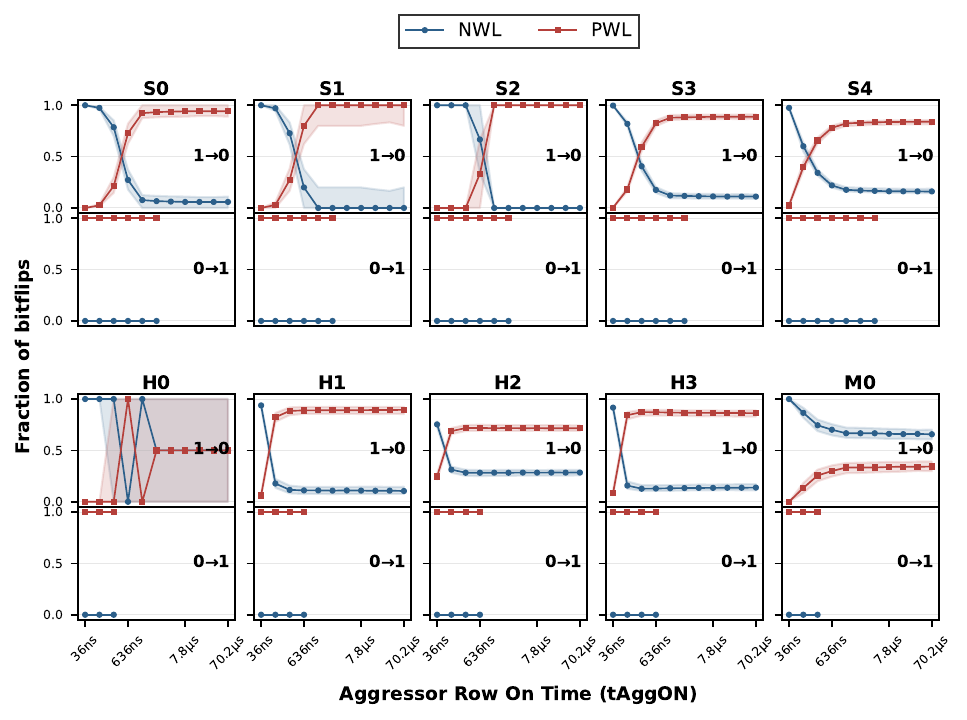}
    \caption{Fraction of 1-to-0 and 0-to-1 bitflips caused by NWL and PWL RowPress.}
    \label{fig:rp-nwlpwl}
    \Description{Fraction of 1-to-0 and 0-to-1 bitflips caused by NWL and PWL RowPress.}
\end{figure}

Figure~\ref{fig:rp-nwlpwl} shows the fraction of 1-to-0 and 0-to-1 bitflips caused by NWL and PWL RowPress across all tested DRAM modules using the NWL and PWL reverse engineering results from Section~\ref{sec:ss-rh-characteristics}. We find that almost all tested DRAM modules show that single-sided PWL RowPress induces 1-to-0 bitflips, while no single-sided NWL RowPress bitflips are observable for large tAggON values where RowPress is dominant in normal DRAM operating conditions.

By comparing the RowPress bitflip locations to the RowHammer bitflip locations, we find that the existing device-level modeling that PWL RowPress induces 1-to-0 bitflips is consistent with experimental characterization. However, we do not observe the NWL RowPress induced 0-to-1 bitflips.

\matchingmechanism{Single-Sided PWL RowPress induces 1-to-0 bitflips.}
\unmatchingmechanism{No single-sided NWL RowPress bitflips are observable in normal DRAM operating conditions.}

\begin{figure}[ht]
    \centering
    \includegraphics[width=\linewidth]{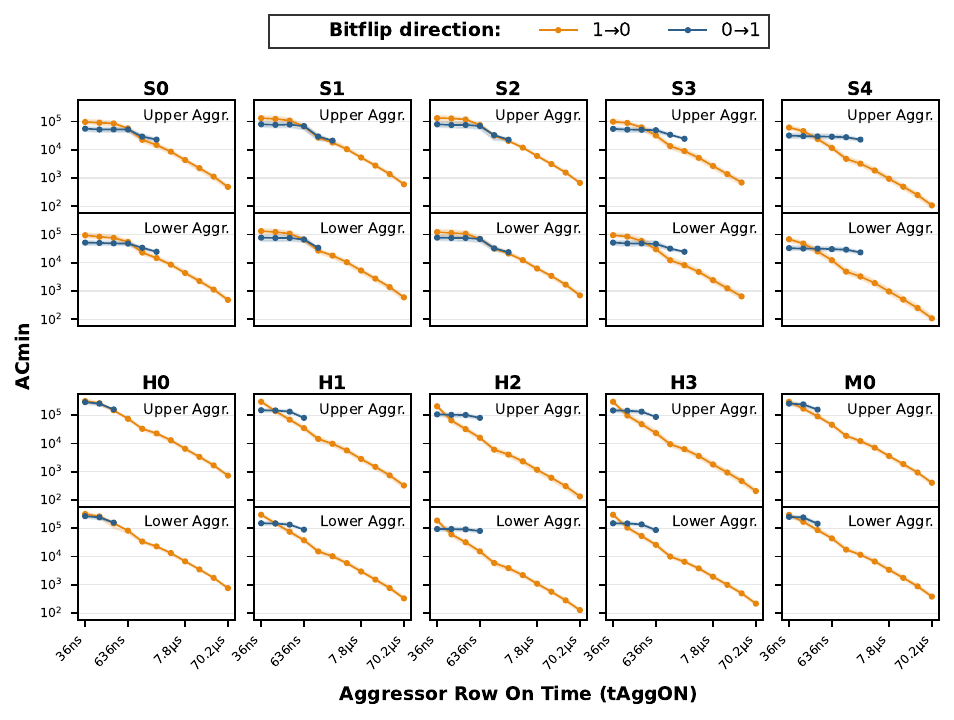}
    \caption{How $\mathrm{AC}_{min}$ changes as tAggON increases for 1-to-0 (yellow) and 0-to-1 (blue) bitflips with the upper (top subplots) and lower (bottom subplots) aggressor row.}
    \label{fig:rp-acmin}
    \Description{How $\mathrm{AC}_{min}$ changes as tAggON increases for 1-to-0 (yellow) and 0-to-1 (blue) bitflips with the upper (top subplots) and lower (bottom subplots) aggressor row.}
\end{figure}

Figure~\ref{fig:rp-acmin} shows how the $\mathrm{AC}_{min}$ of 1-to-0 and 0-to-1 bitflips with the upper and lower aggressor row changes as tAggON increases for all the DRAM modules we test. 

We find that for both NWL and PWL RowPress, their $\mathrm{AC}_{min}$ values \emph{consistently} decrease as tAggON increases. This is not consistent with the existing device-level modeling of RowPress that shows competitive 0-to-1 and 1-to-0 leakage mechanisms between RowHammer and RowPress. For example, for PWL, the opening and closing of the wordline draws electrons away from the victim node, causing 0-to-1 bitflips (i.e., RowHammer). This mechanism is not sensitive to tAggON because the charge traps quickly saturate when the aggressor WL is opened, and the actual leakage happens when the aggressor is closed. When we keep PWL open for a long period of time (i.e., RowPress), its strong electric field will pull electrons from the substrate towards the victim cell, causing 1-to-0 leakage. Since the RowPress-induced leakage is initially weaker than that of RowHammer, $\mathrm{AC}_{min}$ should initially \emph{increase} instead of decrease as tAggON increases (due to the two opposing leakage canceling out each other) according to existing device-level modeling. 
\unmatchingmechanism{As tAggON increases, the $\mathrm{AC}_{min}$ of both 1-to-0 and 0-to-1 bitflips consistently decreases. However, according to device-level modeling, due to the competitive relationship between the leakage mechanisms of RowHammer and RowPress, $\mathrm{AC}_{min}$ should initially increase until RowPress suppresses RowHammer.}

\section{Updated Device-Level Modeling}
\label{sec:new_device_level}
\subsection{TCAD Simulation Methodology}

{We use Sentaurus TCAD ~\cite{TCAD2018manual} to build DRAM cell structures  and perform device-level simulations to investigate the underlying mechanisms of RowHammer and RowPress. The simulated device adopts a saddle-fin-based recessed channel access transistor (RCAT)~\cite{Park2006SFinRCAT, Chae2024SingleMetalBCAT} with buried wordlines. The dimensions of the DRAM cell structures are modeled based on typical 1z-nm DRAM process technology.}

\begin{figure}[ht]
    \centering
    \includegraphics[width=0.8\linewidth]{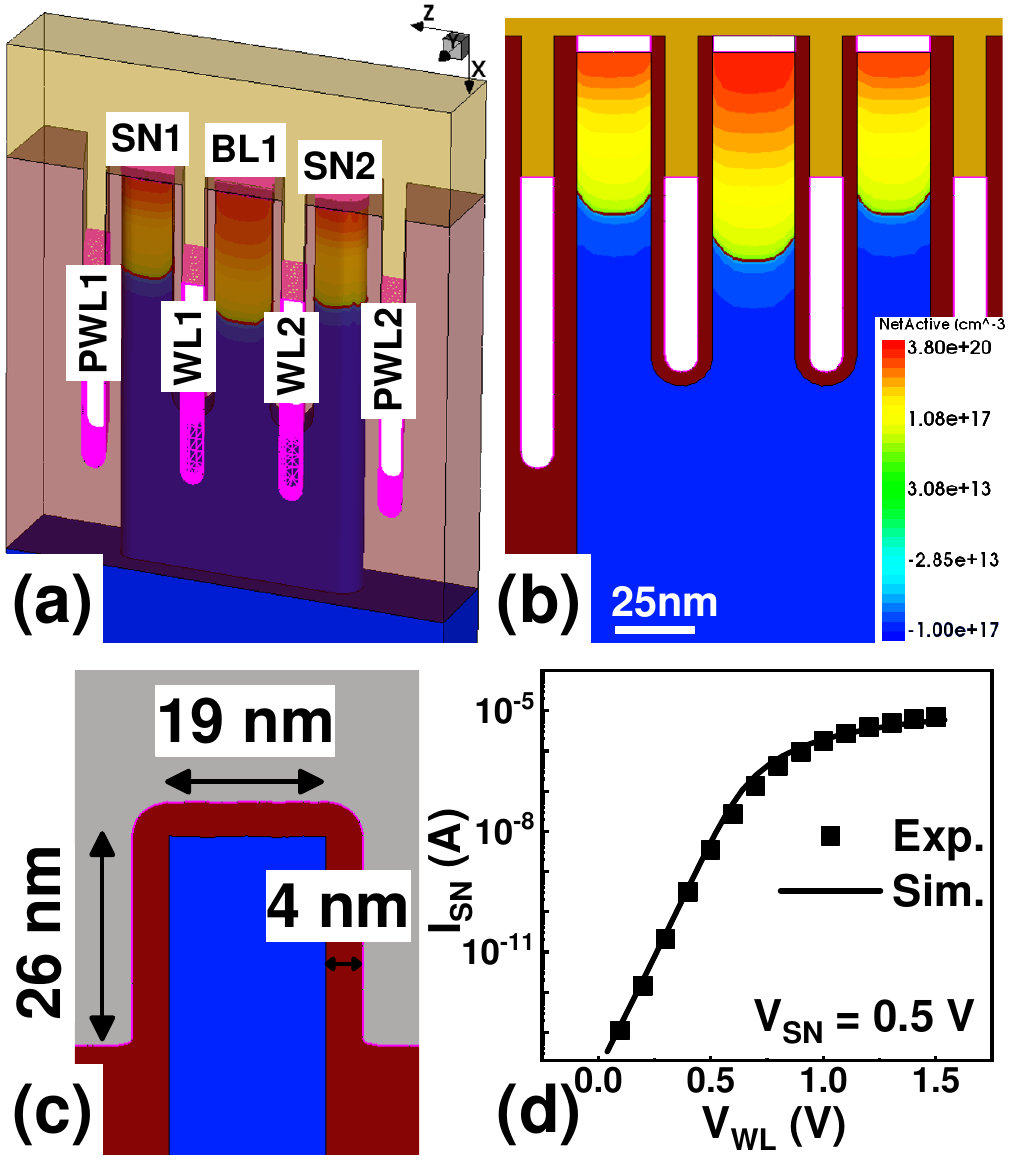}
    \caption{(a) 3D view of the saddle-fin-based DRAM structure highlighting the contact locations. Cross-sectional view cut along (b) active region direction and (c) WL direction, with a depiction of the doping profile. (d) $I_\mathrm{SN}$-$V_\mathrm{WL}$ calibration using experimental data~\cite{Yang2013SuperiorImprovements} from a DRAM manufacturer.}
    \label{fig:tcad-structure}
    \Description{3D view and cross-sections of the saddle-fin-based DRAM structure with doping profile and ISN-VWL calibration.}
\end{figure}

{Figure~\ref{fig:tcad-structure}(a) shows the modeled DRAM cell structure. Following the same convention as in Section~\ref{sec:dram_layout_background}, we define WL1 as the victim WL (i.e., SN1 is the victim cell storage node). WL1's neighboring WL that shares the active region (i.e., WL2) is therefore the NWL aggressor. PWL1 serves as the passing WL (PWL) aggressor.} The cross-sectional views {of the modeled DRAM structure} along the active region and WL directions, including doping profiles and critical dimensions, are shown in Figure~\ref{fig:tcad-structure}(b) and (c), {respectively}.

\noindent\textbf{Physical Models and Calibration.}
The device simulation employs the following physical models: inversion and accumulation layer mobility, high-field saturation, bandgap narrowing, and Shockley-Read-Hall (SRH) recombination with doping dependence, temperature dependence, and field enhancement. In addition, the Hurkx band-to-band tunneling model and the e-quantum correction model were utilized. Figure~\ref{fig:tcad-structure}(d) shows the simulated $I_\mathrm{SN}$-$V_\mathrm{WL}$ characteristics calibrated with experimental data~\cite{Yang2013SuperiorImprovements} from a DRAM manufacturer, ensuring the accuracy of the device model. The baseline I-V characteristics shown in Figure~\ref{fig:tcad-structure} do not include trap specifications and serve as a reference for subsequent simulations.

\begin{figure}[ht]
    \centering
    \includegraphics[width=0.85\linewidth]{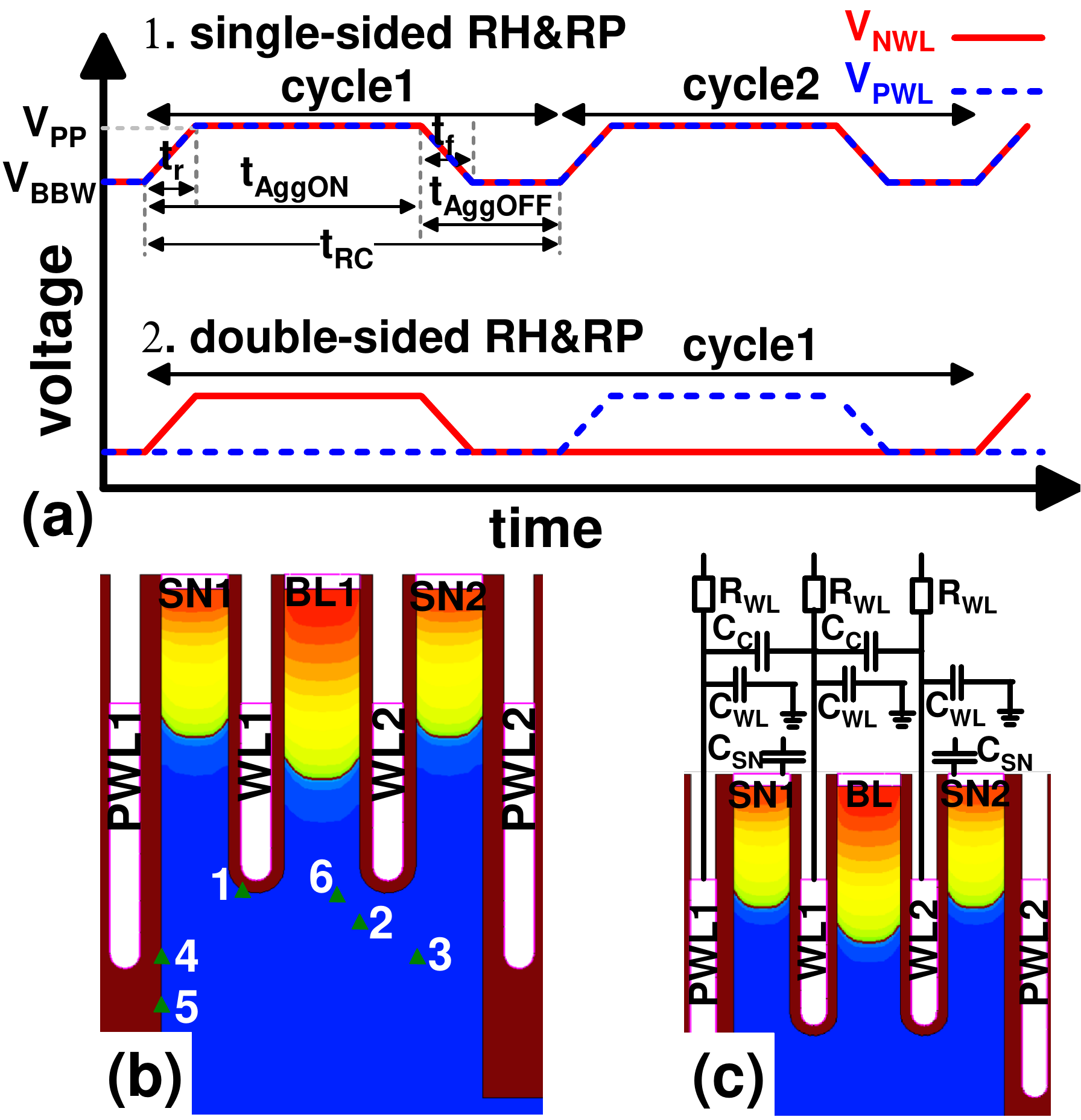}
    \caption{(a) Aggressor WL waveforms for single-sided and double-sided RowHammer and RowPress simulations. (b) Cross-sectional view of the active region, where triangles indicate the locations of single interface e-traps. (c) Mixed-mode simulation circuitry for CC analysis, including WL resistance ($R_\mathrm{WL}$), WL-to-ground capacitance ($C_\mathrm{WL}$), and inter-WL coupling capacitance ($C_\mathrm{C}$).}
    \label{fig:tcad-methodology}
    \Description{Aggressor WL waveforms, interface trap locations, and mixed-mode simulation circuitry for capacitive crosstalk analysis.}
\end{figure}

\noindent\textbf{Trap-Assisted Electron Migration (EM) Modeling.}
To investigate trap-assisted leakage mechanisms, {we place interface electron traps (e-traps) at the silicon-oxide interface. In this study, we adopt a single-trap configuration to isolate the impact of trap location on electron migration behavior.} Figure~\ref{fig:tcad-methodology}(b) illustrates the positions of single interface e-traps in the active region cross-section. Three categories of trap locations are defined: traps near the aggressor NWL (e.g., trap~2, trap~3, and trap~6), traps near the aggressor PWL (e.g., trap~4 and trap~5), and a trap near the victim WL1 (e.g., trap~1). {Such a categorization} allows systematic evaluation of the sensitivity of leakage behavior to trap location and identification of e-trap-sensitive regions for different types of disturbances. While the single-trap configuration is used for sensitivity analysis, its spatial influence is inherently limited. Therefore, a spatially uniform interface trap density ($D_\mathrm{IT}$) is adopted in Section~5.3 to extend the affected region and capture the collective impact of traps.

\noindent\textbf{Capacitive Crosstalk (CC) Modeling.}
{We investigate the capacitive crosstalk (CC) mechanism using mixed-mode TCAD simulation as shown in Figure~\ref{fig:tcad-methodology}(c).} {The simulation incorporates} WL resistance ($R_\mathrm{WL}$), WL-to-ground capacitances ($C_\mathrm{WL}$), and coupling capacitances ($C_\mathrm{C}$) between adjacent WLs to generate a crosstalk noise voltage ($V_\mathrm{noise}$) on the victim WL~\cite{Zhou2024Unveiling, Zhou2024Understanding, Jie2024Understanding}. The CC-induced leakage enhancement effect is attributed to increased subthreshold leakage caused by the additional $V_\mathrm{noise}$ on WL1~\cite{walker2021ondramrowhammer}. A higher peak value and extended pulse width of $V_\mathrm{noise}$ lead to increased leakage.

\noindent\textbf{RowHammer and RowPress Simulation Flow.}
{We evaluate the RowHammer and RowPress effects using TCAD mixed-mode transient simulations.} Figure~\ref{fig:tcad-methodology}(a) shows the aggressor WL waveforms for single-sided and double-sided access patterns. In the single-sided case, only one adjacent WL is activated, while the other remains inactive and is held at $V_\mathrm{BBW}$. For single-sided patterns, the activation count equals the cycle number (doubled for double-sided). 

Table~\ref{tab:sim-params} summarizes the nominal simulation parameters. The $t_\mathrm{AggOFF}$ is kept constant across all simulations, and the ground plane of the storage capacitor is set to $0.5V_\mathrm{CORE}$. {To study and analyze 1-to-0 failures, the victim storage node (SN1) is initialized to $V_\mathrm{CORE}$, and failures happen when SN1 voltage decreases from $V_\mathrm{CORE}$ by more than a defined failure threshold (0.1~V in this study).} {To study and analyze 0-to-1 failures, SN1 is initialized to 0~V, and failures happen when SN1 voltage increases from 0~V by more than a defined failure threshold (0.1~V in this study).} The number of activations required to reach this threshold is defined as the number of activations to failure ($\mathrm{AC}_{min}$). Unless otherwise specified, simulations are performed within a single active region at 300~K, with the data value of SN2 set to be the opposite of SN1 and $V_\mathrm{BL}$ at $0.5V_\mathrm{CORE}$ for simplicity.

\begin{table}[htbp]
\caption{Nominal Simulation Parameters}
\label{tab:sim-params}
\centering
\resizebox{\columnwidth}{!}{%
\begin{tabular}{@{}cc|cc|cc@{}}
\toprule
\textbf{Parameter} & \textbf{Value} & \textbf{Parameter} & \textbf{Value} & \textbf{Parameter} & \textbf{Value} \\ \midrule
$t_\mathrm{r}$/$t_\mathrm{f}$ & 1.25~ns & $V_\mathrm{BL}$ & 0.5~V & $V_\mathrm{CORE}$ & 1~V \\
$t_\mathrm{AggON}$ & 35~ns & $V_\mathrm{SUB}$ & $-$0.7~V & $C_\mathrm{C}$ & 515~fF \\
$t_\mathrm{AggOFF}$ & 13.75~ns & $V_\mathrm{WL1}$ & $-$0.2~V & $C_\mathrm{WL}$ & 740~fF \\
$V_\mathrm{PP}$ & 2.5~V & $V_\mathrm{PWL2}$ & $-$0.2~V & $R_\mathrm{WL}$ & 1~k$\Omega$ \\
$V_\mathrm{BBW}$ & $-$0.2~V & WL WF & 4.5~eV & $C_\mathrm{SN}$ & 10~fF \\ \bottomrule
\end{tabular}%
}
\end{table}

\subsection{Bitflip Direction of Double-Sided RowHammer}
\begin{figure}[ht]
    \centering
    \includegraphics[width=\linewidth]{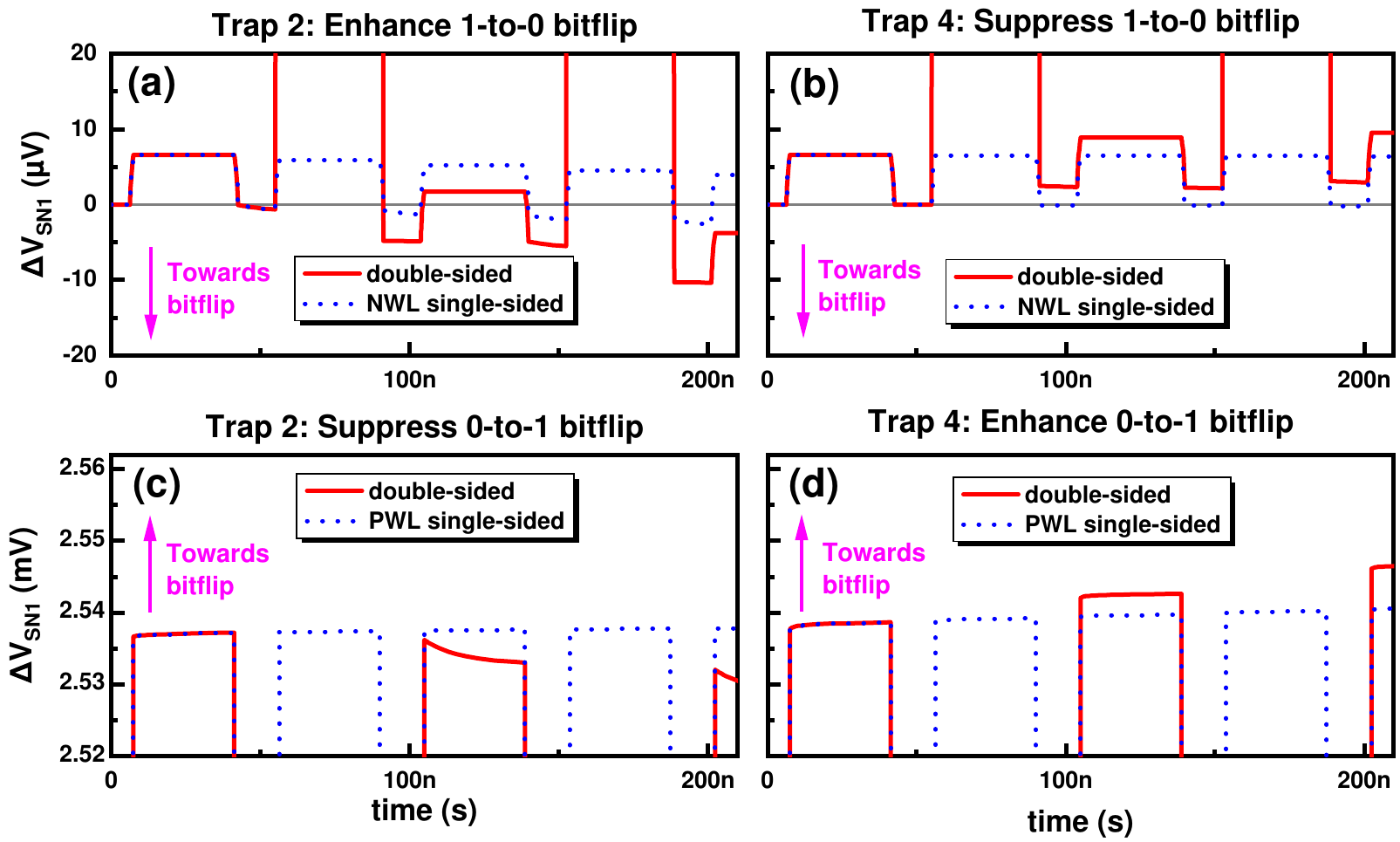}
    \caption{Impact of different charge trap locations on the change in victim storage node voltage ($\Delta V_\mathrm{SN1}$) of double-sided RowHammer (solid red lines) compared to single-sided RowHammer (dashed blue lines). (a)-(b): $\Delta V_\mathrm{SN1}$ when the victim cell initially stores a 1 (i.e., $\Delta V_\mathrm{SN1} = V_\mathrm{SN1} - V_\mathrm{CORE}$) when the charge trap is near the NWL (trap 2, a) and the PWL (trap 4, b). (c)-(d): $\Delta V_\mathrm{SN1}$ when the victim cell initially stores a 0 (i.e., $\Delta V_\mathrm{SN1} = V_\mathrm{SN1} - \mathrm{0V}$) when the charge trap is near the NWL (trap 2, c) and the PWL (trap 4, d).}
    \label{fig:ds-rh-bfdir-1}
    \Description{Impact of different charge trap locations on the change in victim storage node voltage ($\Delta V_\mathrm{SN1}$) of double-sided RowHammer (solid red lines) compared to single-sided RowHammer (dashed blue lines).}

\end{figure}

{We find that the bitflip direction of double-sided RowHammer is sensitive to the trap location. Figure~\ref{fig:ds-rh-bfdir-1} shows the change in victim storage node voltage $\Delta V_\mathrm{SN1}$ (y-axis, $\Delta V_\mathrm{SN1}=V_\mathrm{SN1}-V_\mathrm{CORE}$ for 1-to-0 bitflips, and $\Delta V_\mathrm{SN1}=V_\mathrm{SN1}-\mathrm{0V}$ for 0-to-1 bitflips) for trap location near the NWL (i.e., trap location 2 in Figure~\ref{fig:tcad-methodology}.b) and near the PWL (i.e., trap location 4 in Figure~\ref{fig:tcad-methodology}.b). The solid red lines shows the $\Delta V_\mathrm{SN1}$ of double-sided RowHammer, compared to single-sided RowHammer in dashed blue lines (NWL for 1-to-0 bitflips, and PWL for 0-to-1 bitflips).}

We observe that for a given victim cell with a fixed charge trap configuration, double-sided RowHammer enhances either 1-to-0 or 0-to-1 bitflips while suppressing the other, depending on the location of the charge trap, compared to single-sided RowHammer. Figure~\ref{fig:ds-rh-bfdir-1}(a) shows that, for a victim cell initially storing a 1 (i.e., $V_\mathrm{SN1}$ is initially $V_\mathrm{CORE}$), if the charge trap is located near the NWL (i.e., location 2), double-sided RowHammer (solid red lines) reduces $V_\mathrm{SN1}$ more than single-sided RowHammer (dashed blue lines), enhancing 1-to-0 bitflips. Figure~\ref{fig:ds-rh-bfdir-1}(b) shows that, for a victim cell initially storing a 1, if the charge trap is located near the PWL (i.e., location 4), double-sided RowHammer \emph{increases} $V_\mathrm{SN1}$, suppressing 1-to-0 bitflips. Figure~\ref{fig:ds-rh-bfdir-1}(c) and (d) show that, for a victim cell initially storing a 0 (i.e., $V_\mathrm{SN1}$ is initially $0\mathrm{V}$), 1) charge trap near the NWL does not keep increasing $V_\mathrm{SN1}$ for double-sided RowHammer unlike single-sided RowHammer, suppressing 0-to-1 bitflips, and 2) charge trap near the PWL increases $V_\mathrm{SN1}$ more than single-sided RowHammer, enhancing 0-to-1 bitflips.

\begin{figure}[ht]
    \centering
    \includegraphics[width=0.95\linewidth]{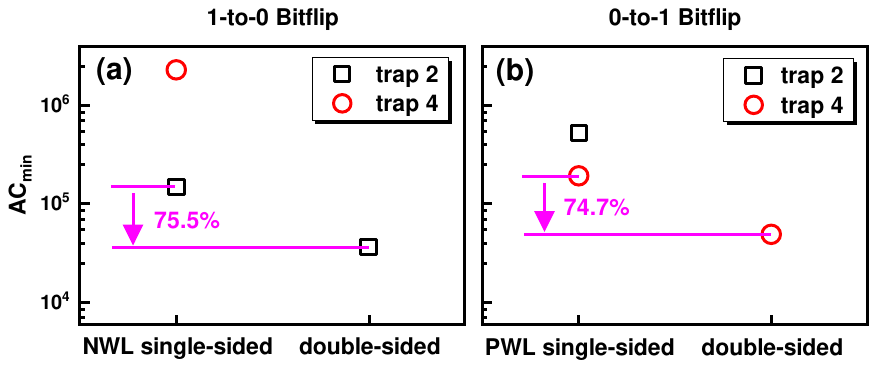}
    \caption{$\mathrm{AC}_{min}$ of 1-to-0 and 0-to-1 bitflips for different charge trap locations at the temperature of 325K; double-sided RowHammer compared to single-Sided RowHammer.}
    \label{fig:ds-rh-bfdir-2}
    \Description{$\mathrm{AC}_{min}$ of 1-to-0 and 0-to-1 bitflips for different charge trap locations.}
\end{figure}

{Figure~\ref{fig:ds-rh-bfdir-2}(a) and (b) compare the $\mathrm{AC}_{min}$ of 1-to-0 and 0-to-1 bitflips under single- and double-sided RowHammer, respectively, for trap locations 2 (black square) and 4 (red circle). We observe that when the trap is at location 2 (near NWL), 1) the $\mathrm{AC}_{min}$ of 1-to-0 bitflip for double-sided RowHammer reduces by 75.5\% compared to single-sided RowHammer (NWL), and 2) we do not observe double-sided RowHammer 0-to-1 bitflip. When the trap is at location 4 (near PWL), 1) the $\mathrm{AC}_{min}$ of 0-to-1 bitflip for double-sided RowHammer reduces by 74.7\% compared to single-sided RowHammer (PWL), and 2) we do not observe double-sided RowHammer 1-to-0 bitflip.}
 
\begin{figure}[ht]
    \centering
    \includegraphics[width=\linewidth]{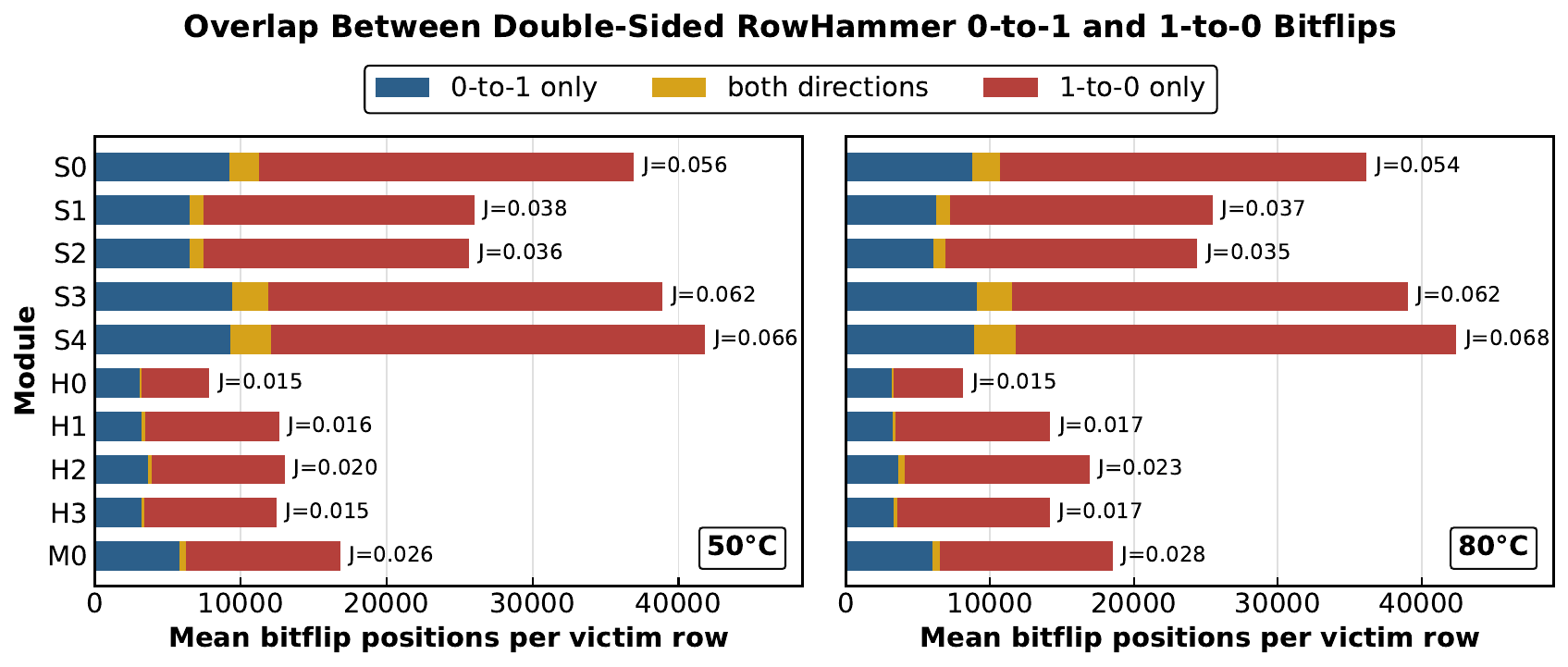}
    \caption{Overlap between 0-to-1 and 1-to-0 bitflip positions for double-sided RowHammer.}
    \label{fig:ds-bf-overlap}
    \Description{Overlap between 0-to-1 and 1-to-0 bitflip positions for double-sided RowHammer.}
\end{figure}

To further validate the device-level mechanism that the charge trap locations determine the bitflip direction of double-sided RowHammer, we analyze the overlap of 0-to-1 and 1-to-0 bitflips when we hammer both aggressor rows 1.28M times. Figure~\ref{fig:ds-bf-overlap} shows the average number of observed bitflip positions per victim row that only shows 0-to-1 (blue) or 1-to-0 (red) bitflips and those show both bitflip directions (yellow), at temperatures 50$^\circ$C and 80$^\circ$C. For each module, we annotate the Jaccard index, defined as the ratio between the number of observed bitflip positions that exhibit both bitflip directions and the union of positions that exhibit either direction.

We observe that for all tested modules, only a small fraction of victim cells experience both 1-to-0 and 0-to-1 bitflips. For example, across all tested modules, the mean Jaccard overlap only ranges from 1.49\% to 6.77\%. Even for module S4, which exhibits the largest overlap, only 6.65\% and 6.77\% of observed bitflip positions show both directions at 50$^\circ$C and 80$^\circ$C, respectively. For module M0, the overlap is only 2.57\% at 50$^\circ$C and 2.77\% at 80$^\circ$C. Such small overlaps between the 0-to-1 and 1-to-0 double-sided RowHammer bitflips supports the device-level mechanism that the position of the charge traps determine the bitflip direction of double-sided RowHammer.

 \takeaway{The bitflip direction of double-sided RowHammer is mainly determined by the charge trap location. Charge traps near the NWL enhance 1-to-0 and suppress 0-to-1 bitflips. Charge traps near the PWL enhance 0-to-1 and suppresses 1-to-0 bitflips.}

\subsection{$\mathrm{AC}_{min}$ Difference Between 0-to-1 and 1-to-0 Bitflips of Double-Sided RowHammer}

We find that both 1) trap-assisted electron migration (EM) and 2) capacitive crosstalk (CC) can explain the real-chip characterization result where the $\mathrm{AC}_{min}$ of 0-to-1 bitflips is smaller than that of 1-to-0 bitflips.

\begin{figure}[ht]
    \centering
    \includegraphics[width=\linewidth]{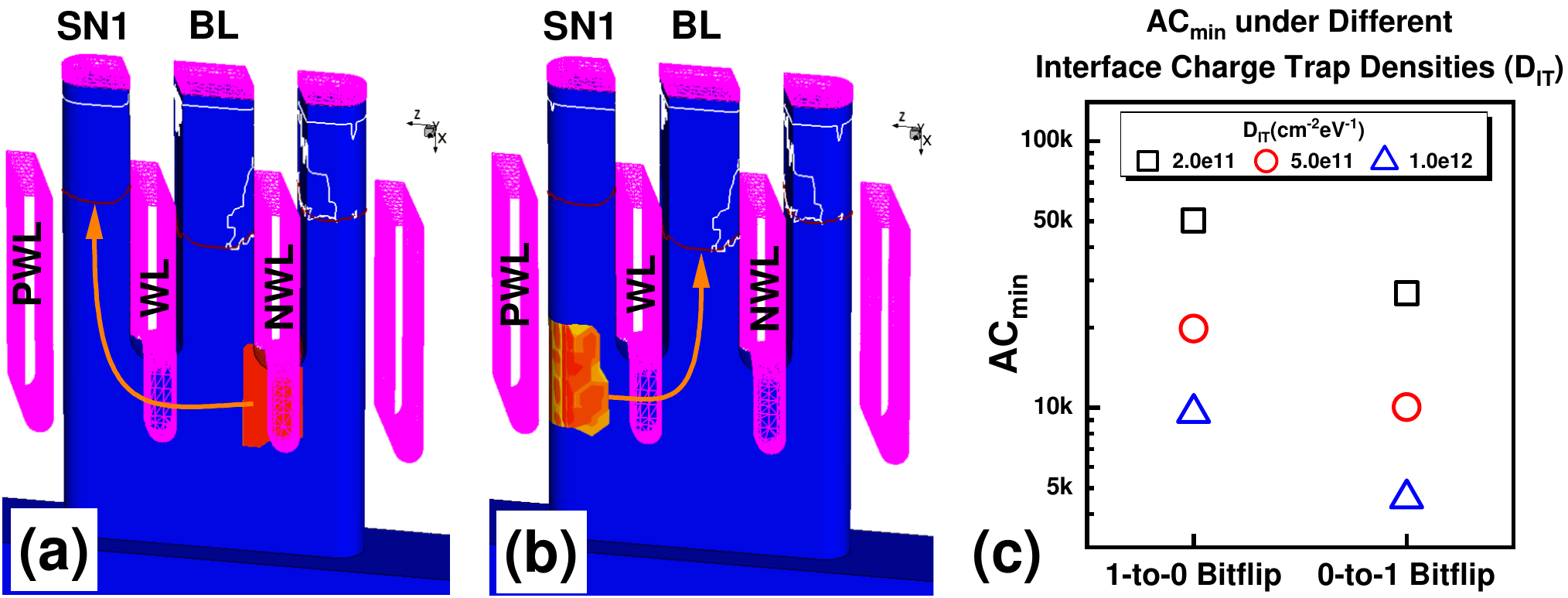}
    \caption{Interface trap regions used for simulating double-sided RowHammer and the electron migration paths for (a) 1-to-0 bitflips and (b) 0-to-1 bitflips. (c) $\mathrm{AC}_{min}$ of 1-to-0 and 0-to-1 bitflips under double-sided RowHammer at different interface trap densities $D_\mathrm{IT}$, using the same trap region area.}
    \label{fig:ds-rh-trap-regions}
    \Description{Interface trap regions and NAC comparison for 1 and 0 failures under double-sided RowHammer at different interface trap densities.}
\end{figure}

\noindent\textbf{Trap-Assisted EM Perspective.}
We find that the electron migration efficiency for the 0-to-1 bitflips is higher than that of 1-to-0 bitflips. We observe that compared to the NWL, the PWL is physically closer to the victim storage node, resulting in a stronger modulation of the local electric field around the SN during activation and pre-charge operations. As a result, the PWL has a more direct impact on electron transport associated with the 0-to-1 bitflips. Figure~\ref{fig:ds-rh-trap-regions}(a) and (b) shows the structure of the simulated DRAM device with the interface trap regions (red-orange area in the blue substrate) and the electron migration path (orange arrows) for (a) 1-to-0 and (b) 0-to-1 bitflips. The interface trap region is near the NWL (PWL) for 1-to-0 (0-to-1) bitflips. For 1-to-0 bitflips under double-sided RowHammer, electrons primarily migrate from the NWL side toward the victim SN, involving a relatively longer migration path. In contrast, for 0-to-1 bitflips, the net electron loss from the SN is dominated by electron emission when the PWL is activated, followed by migration toward the BL. This process involves a shorter migration path, leading to higher electron migration efficiency. 

To further validate electron migration efficiency mechanism, we perform trap-location-dependent $\mathrm{AC}_{min}$ simulations for both 1-to-0 and 0-to-1 bitflips using the same trap region area and interface charge trap densities ($D_\mathrm{IT}$) as shown in Figure~\ref{fig:ds-rh-trap-regions}(c). We observe that the $\mathrm{AC}_{min}$ of 0-to-1 bitflips is consistently smaller than that for 1-to-0 bitflips under double-sided RowHammer, regardless of the assumed $D_\mathrm{IT}$. This result confirms that the higher migration efficiency associated with the shorter path in the 0-to-1 bitflip case leads to smaller $\mathrm{AC}_{min}$. In addition, other process parameters may also contribute to this difference. For example, increasing the BL junction depth may suppress 1-to-0 bitflips while enhancing 0-to-1 bitflips~\cite{zhou2023double}.

 \takeaway{Compared to 1-to-0 bitflips, 0-to-1 bitflips under double-sided RowHammer requires fewer aggressor row activations due to higher electron migration efficiency associated with the shorter electron migration path to the victim node from near the PWL compared to the NWL.}

\begin{figure}[ht]
    \centering
    \includegraphics[width=\linewidth]{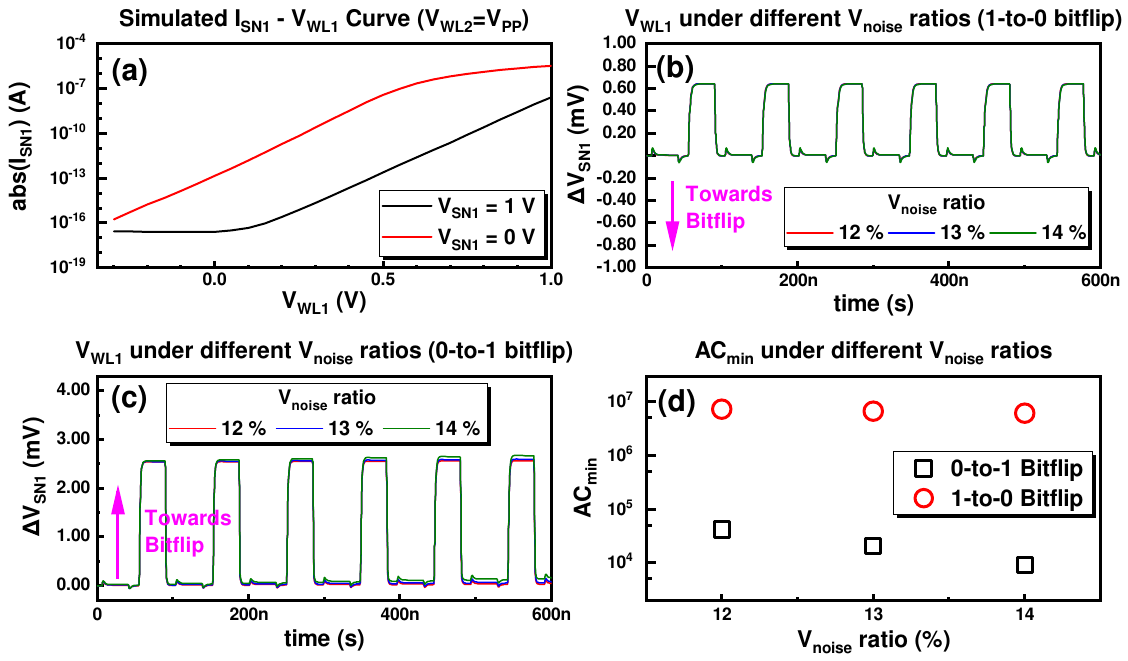}
    \caption{Analysis of CC-induced leakage under double-sided RowHammer. (a) Simulated $I_\mathrm{SN}$-$V_\mathrm{WL}$ characteristics for 0-to-1 (i.e., $V_\mathrm{SN1} = 0\mathrm{V}$) and 1-to-0 (i.e., $V_\mathrm{SN1} = V_\mathrm{CORE} = 1\mathrm{V}$) bitflip conditions. (b) $V_\mathrm{SN1}$ time evolution for 1-to-0 bitflip under different $V_\mathrm{noise}$ ratios, showing weak sensitivity to CC-induced $V_\mathrm{WL}$ increase. (c) $V_\mathrm{SN1}$ time evolution for 0-to-1 bitflip under different $V_\mathrm{noise}$ ratios, showing enhanced leakage with increasing $V_\mathrm{noise}$. (d) Corresponding $\mathrm{AC}_{min}$ comparison for 0-to-1 and 1-to-0 bitflips as a function of $V_\mathrm{noise}$ ratio, indicating consistently smaller $\mathrm{AC}_{min}$ for 0-to-1 bitflip.}
    \label{fig:ds-rh-cc}
    \Description{Analysis of capacitive crosstalk induced leakage under double-sided RowHammer showing 0-to-1 bitflip is more severe than 1-to-0 bitflip.}
\end{figure}

\noindent\textbf{Capacitive Crosstalk (CC) Perspective.}
We also find that capacitive crosstalk can enhance 0-to-1 bitflips under double-sided RowHammer because the effective threshold voltage of the access transistor under the 0-to-1 bitflip condition is \emph{lower} compared to that of 1-to-0 bitflip. Figure~\ref{fig:ds-rh-cc} shows the analysis of the CC-induced leakage behavior under double-sided RowHammer, including the (a) leakage current, (b)-(c) the change in victim storage node voltage, and (d) $\mathrm{AC}_{min}$. The simulations are configured without the impact of traps. 

Figure~\ref{fig:ds-rh-cc}(a) shows the simulated $I_\mathrm{SN}$-$V_\mathrm{WL}$ characteristics for both 0-to-1 and 1-to-0 bitflip conditions. {We observe} that the effective threshold voltage {of the access transistor} under the 1-to-0 bitflip condition is higher than that under the 0-to-1 bitflip condition, indicating that a larger $V_\mathrm{WL}$ is required to significantly increase $I_\mathrm{SN}$ in the 1-to-0 bitflip case. Under the CC mechanism, when the aggressor WL is activated, the coupling-induced noise results in a transient increase in $V_\mathrm{WL1}$. For example, with $V_\mathrm{PP} = 2.5$~V and $V_\mathrm{noise\_ratio} = 12\%$, $V_\mathrm{WL1}$ increases from $-0.2$~V to $0.124$~V. This increase in $V_\mathrm{WL1}$ strongly enhances $I_\mathrm{SN}$ under the 0-to-1 bitflip condition, while it has a negligible impact under the 1-to-0 bitflip condition due to the higher effective threshold voltage. 

{Figure~\ref{fig:ds-rh-cc}(b) and (c) further confirms that 0-to-1 bitflip is more sensitive to the CC-induced $V_\mathrm{WL}$ increase.} The SN1 voltage evolution under 1-to-0 bitflip shows little dependence on the noise amplitude, whereas under 0-to-1 bitflip, the leakage becomes more pronounced as the $V_\mathrm{noise}$ ratio increases. As a result, the charge loss in the 0-to-1 bitflip case accumulates more rapidly. Figure~\ref{fig:ds-rh-cc}(d) summarizes the corresponding $\mathrm{AC}_{min}$ results. It is observed that the $\mathrm{AC}_{min}$ for 0-to-1 bitflip is consistently smaller than that for 1-to-0 bitflip across different $V_\mathrm{noise}$ ratios. 

 \takeaway{When capacitive crosstalk is strong, it enhances the 0-to-1 bitflip much more than the 1-to-0 bitflip under double-sided RowHammer because the effective threshold voltage of the DRAM access transistor under the 1-to-0 bitflip condition is higher than that under the 0-to-1 bitflip condition.}

\subsection{Difference in the Number of 0-to-1 and 1-to-0 Bitflips in Double-Sided RowHammer}

The observed higher occurrence of 1-to-0 bitflips than 0-to-1 bitflips at sufficiently large AC is a statistical result obtained from a large population of cells in real DRAM chips, and quantitatively identifying its exact root cause is beyond the scope of this work. However, prior process and device studies suggest a plausible qualitative explanation from the perspective of DRAM structure and fabrication. The asymmetry of the number of 1-to-0 bitflips than 0-to-1 bitflips is likely related to the high process sensitivity of WL-related fabrication in advanced DRAM that uses the buried channel array transistor (BCAT) structure~\cite{Yang2013SuperiorImprovements, Chae2024SingleMetalBCAT}. Prior studies have shown that WL profile control is highly sensitive to the local geometry and material environment, and that the active-area region is more vulnerable to profile non-uniformity than the shallow trench isolation (STI) region~\cite{Jang2025Gateoxidetechnology}. BCAT process studies have further shown that local electrical variability is strongly influenced by active-region dimensions and profile~\cite{Jeon2017InvestigationOnTheLocal}. In this context, the saddle-fin region adjacent to the NWL is formed over the active area with a more complex local topology, whereas the PWL-side region mainly overlies STI. Such differences in local geometry and material environment make the NWL-side region more susceptible to process-induced profile non-uniformity and interfacial degradation than the PWL-side region.

Therefore, although a direct experimental comparison of interface-trap density between the NWL-adjacent saddle-fin region and the PWL-side region has not yet been reported, it is physically reasonable to infer that the WL patterning and etching process may lead to a higher local interface-trap density in the saddle-fin region adjacent to the NWL than in the sidewall region near the PWL. As a result, for a large population of DRAM cells, traps are more likely to reside in NWL-sensitive regions, thereby increasing the likelihood of 1-to-0 bitflips. This asymmetry in trap distribution can therefore lead to a higher occurrence of 1-to-0 bitflips than 0-to-1 bitflips under double-sided RowHammer when the hammer count is sufficiently large.

 \takeaway{The current DRAM structure and manufacturing process statistically create favorable conditions for 1-to-0 bitflips over 0-to-1 bitflips under double-sided RowHammer.}

\subsection{The Missing NWL-Induced RowPress 0-to-1 Bitflips}

Previous simulation studies {suggest} the possibility of 0-to-1 bitflips induced by NWL RowPress, where opening the NWL increases the leakage current from the storage node~\cite{Zhou2024Unveiling, Zhou2024Understanding}. However, whether such 0-to-1 bitflips can be experimentally observed depends on whether the induced leakage is sufficiently large to cause significant charge loss within the retention time window (e.g., 64~ms). {To directly evaluate the experimental observability of this behavior at the transistor level, we perform wafer-level electrical measurements of the leakage current (i.e., $I_\mathrm{SN1}$, defined as current flowing \emph{away} from the victim storage node) on real \emph{fabricated} test structures consisting of sub-20~nm industry-grade DRAM cell access transistors in addition to the TCAD analysis. Since the relevant leakage current is extremely small at the single-cell level, the test structure consists of multiple cell transistors connected in parallel to obtain a reliably measurable current.} 

\begin{figure}[ht]
    \centering
    \includegraphics[width=\linewidth]{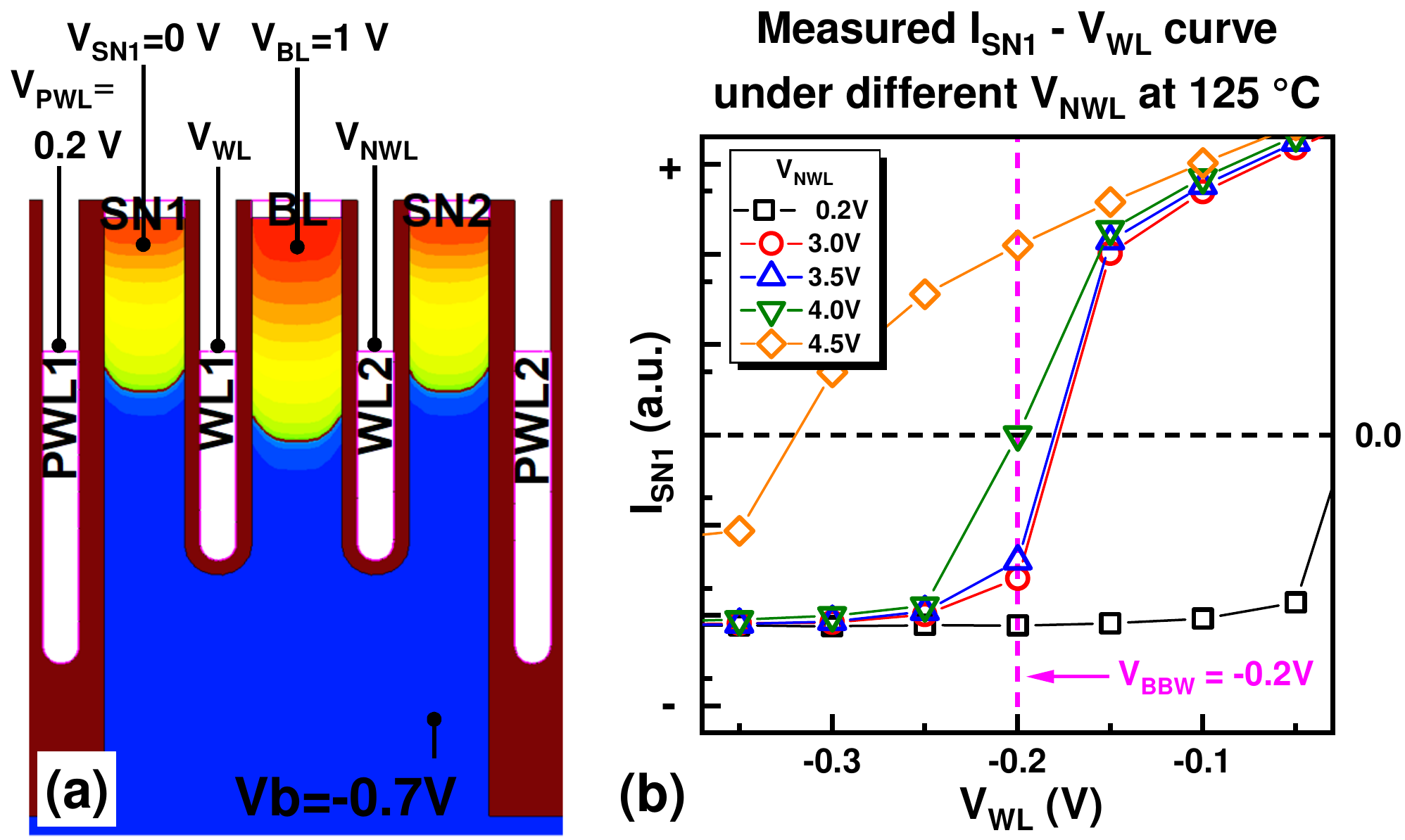}
    \caption{(a) Device schematic and bias conditions used to characterize NWL RowPress-induced 0-to-1 bitflips. (b) Measured $I_\mathrm{SN1}$-$V_\mathrm{WL}$ characteristics under 0-to-1 bitflip bias conditions at different $V_\mathrm{NWL}$ from real fabricated test devices.}
    \label{fig:awl-rp-measurement}
    \Description{Device schematic and measured ISN1-VWL characteristics for NWL RowPress-induced 0 bitflip at different NWL voltages.}
\end{figure}

Figure~\ref{fig:awl-rp-measurement}(a) shows the device structure and bias conditions we apply to the test devices. Figure~\ref{fig:awl-rp-measurement}(b) presents the measured $I_\mathrm{SN1}$-$V_\mathrm{WL}$ characteristics under 0-to-1 bitflip bias conditions for different NWL voltages ($V_\mathrm{NWL}$). Note that the y-axis is \emph{negative} $I_\mathrm{SN1}$ to match the flow of electrons, i.e., if a data point is above 0, electrons move \emph{away} from the victim storage node, contributing to 0-to-1 bitflip. We observe that opening the NWL (i.e., $V_\mathrm{NWL} > V_\mathrm{BBW} = -0.2V$, red, blue, green, and orange symbols and lines in Figure~\ref{fig:awl-rp-measurement}(b)) leads to a noticeable increase in $I_\mathrm{SN1}$, which is consistent with simulation results indicating enhanced subthreshold leakage. However, the leakage current relevant to data retention is determined by $I_\mathrm{SN1}$ when the victim wordline is closed (i.e., $V_\mathrm{WL}$=$V_\mathrm{BBW}$=$-0.2$~V), highlighted with the vertical dashed magenta line in Figure~\ref{fig:awl-rp-measurement}(b). When $V_\mathrm{NWL}$ is below 3.5~V, $I_\mathrm{SN1}$ at $V_\mathrm{BBW}$ remains positive, indicating that electron loss from the storage node is \emph{insufficient} to induce a 0-to-1 bitflip. As $V_\mathrm{NWL}$ increases to 4.0~V, $I_\mathrm{SN1}$ at $V_\mathrm{BBW}$ becomes negative, and further increasing $V_\mathrm{NWL}$ to 4.5~V results in a sufficiently large negative current that can trigger 0-to-1 bitflip.

These results indicate that, under typical operating conditions where $V_\mathrm{NWL}$ is limited to $V_\mathrm{PP}$, NWL RowPress does not generate a sufficiently large negative $I_\mathrm{SN1}$ at $V_\mathrm{BBW}$. As a result, even under prolonged hammering, the induced leakage remains too weak to cause observable 0-to-1 bitflip. Thus, direct measurements on fabricated DRAM cell transistors confirm that NWL-induced RowPress leakage is insufficient to overcome the intrinsic data-retention leakage of the victim cell under normal operating conditions.

 \takeaway{Under normal DRAM operating conditions, the NWL-induced RowPress leakage is not sufficiently strong to overcome the data retention leakage in the victim cell, leading to unobservable 0-to-1 bitflip.}

\subsection[ACmin of 1-to-0 and 0-to-1 bitflips both decrease as tAggON increases in RowPress]{$\mathrm{AC}_{min}$ of 1-to-0 and 0-to-1 bitflips both decrease as $t_\mathrm{AggON}$ increases in RowPress}

Previous studies have shown that, for 1-to-0 bitflip, $\mathrm{AC}_{min}$ decreases with increasing $t_\mathrm{AggON}$, which has been well explained by the enhanced leakage induced by longer WL activation time~\cite{zhou2023double}. However, our experimental observations indicate that $\mathrm{AC}_{min}$ for 0-to-1 bitflip also decreases with increasing $t_\mathrm{AggON}$, which appears inconsistent with prior simulation results where, for PWL RowHammer-induced 0-to-1 bitflip, $\mathrm{AC}_{min,\mathrm{0-to-1}}$ increases with increasing $t_\mathrm{AggON}$~\cite{Zhou2024Unveiling, Zhou2024Understanding}. To resolve this discrepancy, {we analyze the $t_\mathrm{AggON}$ dependence of $\mathrm{AC}_{min,\mathrm{0-to-1}}$} from the perspective of the competition between 0-to-1 and 1-to-0 bitflip mechanisms under PWL-induced RowPress. We identify that {the density of hole traps in the bulk region of the device ($D_\mathrm{HT,bulk}$)} is a key factor in modulating the strength of the PWL RowPress-induced 1-to-0 bitflip.

\begin{figure}[ht]
    \centering
    \includegraphics[width=\linewidth]{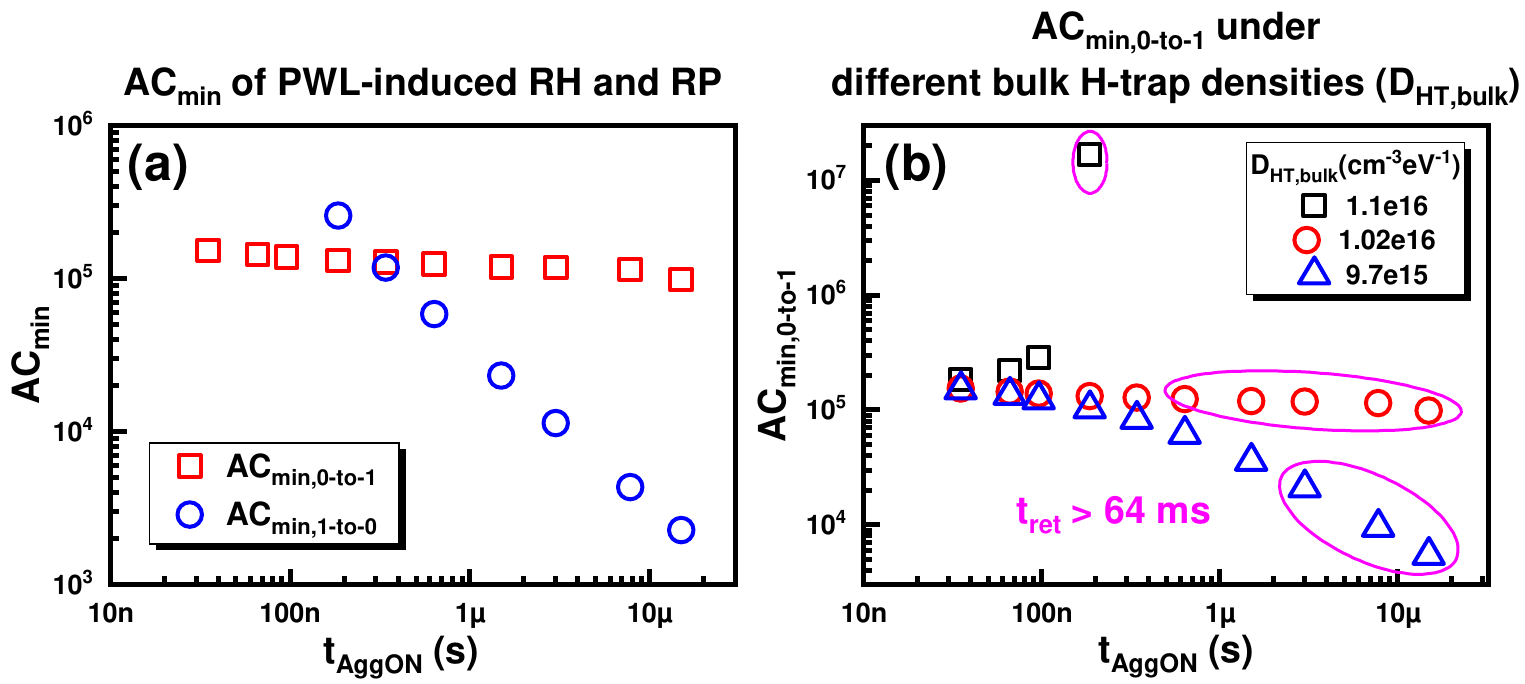}
    \caption{Dependence of $\mathrm{AC}_{min}$ on $t_\mathrm{AggON}$ for PWL-induced RowHammer and RowPress; trap location 4 (near PWL), 0.4eV. (a) {$\mathrm{AC}_{min}$ for 1-to-0 bitflip ($\mathrm{AC}_{min,\mathrm{1-to-0}}$), 0-to-1 bitflip ($\mathrm{AC}_{min,\mathrm{0-to-1}}$)} as a function of $t_\mathrm{AggON}$ under a relatively low bulk h-trap density ($D_\mathrm{HT,bulk} = 1.02 \times 10^{16}$~cm$^{-3}$eV$^{-1}$). (b) $\mathrm{AC}_{min}$ for 0-to-1 bitflip under different bulk h-trap densities ($D_\mathrm{HT,bulk}$). The circled data points indicate the regime where the retention time exceeds 64~ms.}
    \label{fig:nac-taggon}
    \Description{Dependence of NAC on tAggON for PWL-induced RowHammer and RowPress, showing competition between 0 and 1 bitflip modes.}
\end{figure}

 A lower $D_\mathrm{HT,bulk}$ weakens the PWL RowPress-induced 1-to-0 bitflip because it reduces the substrate electron density and thus suppresses the electron flow toward SN1 during the PWL pulse-on phase. As a result, the corresponding charge-loss process responsible for the 1-to-0 transition becomes less effective. In contrast, a higher $D_\mathrm{HT,bulk}$ increases substrate electron density, enhances the electron flow to SN1 during the PWL pulse-on phase, and therefore strengthens the PWL RowPress-induced 1-to-0 bitflip.

Figure~\ref{fig:nac-taggon}(a) shows how $\mathrm{AC}_{min}$ changes for 0-to-1 and 1-to-0 bitflips as $t_\mathrm{AggON}$ increases for PWL-induced RowHammer and RowPress at a relatively low bulk h-trap density ($D_\mathrm{HT,bulk} = 1.02 \times 10^{16}$~cm$^{-3}$eV$^{-1}$). When the PWL RowPress-induced 1-to-0 bitflip is weak (i.e., under lower $D_\mathrm{HT,bulk}$), the leakage is dominated by the 0-to-1 bitflip, and $\mathrm{AC}_{min,\mathrm{0-to-1}}$ decreases with increasing $t_\mathrm{AggON}$. In this case, the observed $\mathrm{AC}_{min}$ follows $\mathrm{AC}_{min,\mathrm{0-to-1}}$ and thus also decreases with $t_\mathrm{AggON}$, as shown in Figure~\ref{fig:nac-taggon}(a), which is consistent with experimental observations. 

Figure~\ref{fig:nac-taggon}(b) shows how $\mathrm{AC}_{min,\mathrm{0-to-1}}$ changes as $t_\mathrm{AggON}$ increases under different $D_\mathrm{HT,bulk}$. We observe that with a higher $D_\mathrm{HT,bulk}$ (i.e., black squares), the PWL RowPress-induced 1-to-0 bitflip is sufficiently strong that it competes with the 0-to-1 bitflip, leading to an increasing trend of $\mathrm{AC}_{min,\mathrm{0-to-1}}$ as $t_\mathrm{AggON}$ increases. When $D_\mathrm{HT,bulk}$ is lower (i.e., blue triangles), it weakens the PWL RowPress-induced 1-to-0 bitflip and amplifies the decreasing trend of $\mathrm{AC}_{min,\mathrm{0-to-1}}$ as $t_\mathrm{AggON}$ increases. Therefore, the experimentally observed decrease of $\mathrm{AC}_{min}$ for 0-to-1 bitflips with increasing $t_\mathrm{AggON}$ can be explained by the insufficient strength of PWL RowPress-induced 1-to-0 bitflip in current devices.

 \takeaway{PWL-induced RowPress 1-to-0 bitflip is sensitive to the density of hole traps in the bulk region of the device ($D_\mathrm{HT,bulk}$). A higher (lower) $D_\mathrm{HT,bulk}$ strengthens (weakens) PWL-induced RowPress 1-to-0 bitflips.}

\section{Summary}
In this section, we summarize the updated device-level mechanisms to explain RowHammer and RowPress bitflips. Table~\ref{tab:device-level-findings} highlights our key device-level findings that bridges the gap and resolves the inconsistencies between prior device-level works and real-chip characterization results.

\begin{table*}[htbp]
\caption{Summary of Device-Level Findings for RowHammer and RowPress Bitflips}
\label{tab:device-level-findings}
\centering
\fontsize{8.5pt}{7.5pt}\selectfont
\resizebox{\textwidth}{!}{%
\begin{tabular}{@{}m{0.32\textwidth}m{0.28\textwidth}m{0.40\textwidth}@{}}
\toprule
\textbf{Real-Chip Observation} & \textbf{Key Device-Level Factor} & \textbf{Device-Level Mechanism / Implication} \\ \toprule

Double-sided RowHammer induces both 1-to-0 and 0-to-1 bitflips with limited spatial overlap.

&
Location of interface traps near the NWL or PWL side.
&
Trap location determines the bitflip direction: NWL(PWL)-side  traps enhance 1-to-0 (0-to-1) bitflips.
\\ \midrule

\multirow[c]{2}{0.32\textwidth}[-1.0\baselineskip]{Double-sided RowHammer $AC_\mathrm{min}$ of 0-to-1 bitflips is smaller than that of 1-to-0 bitflips.}
&
Electron migration path from the NWL/PWL side to victim storage node.
&
0-to-1 bitflips experience a shorter electron migration path, resulting in higher efficiency and thus a smaller $AC_\mathrm{min}$.

\\ \cmidrule(l){2-3}

&
Effective threshold voltage of the access transistor and WL-coupling-induced noise.
&
The lower effective threshold voltage under the 0-to-1 condition makes capacitive-crosstalk-induced noise more effective in enhancing 0-to-1 leakage.

\\ \midrule

More double-sided RowHammer 1-to-0 than 0-to-1 bitflips with a sufficiently high hammer count.
&
Process-induced geometry/material asymmetry between the NWL-adjacent and PWL-adjacent regions.
&
The NWL-adjacent saddle-fin region is more vulnerable to process non-uniformity and interfacial degradation, increasing local trap formation and favoring 1-to-0 bitflips.

\\ \midrule

NWL-induced RowPress does not produce observable 0-to-1 bitflips under normal DRAM operating conditions.
&
Storage-node current polarity and magnitude near $V_\mathrm{WL}=V_\mathrm{BBW}$.
&
Wafer-level measurements show that NWL-induced RowPress leakage is too weak to overcome intrinsic retention leakage, making 0-to-1 bitflips unobservable under normal operation.

\\ \midrule

For PWL-induced RowPress, both 1-to-0 and 0-to-1 $AC_\mathrm{min}$ values decrease as $t_\mathrm{AggON}$ increases.
&
Competition between PWL-induced 1-to-0 and 0-to-1 leakage mechanisms, modulated by bulk hole-trap density.
&
Increasing $t_\mathrm{AggON}$ strengthens PWL-induced leakage, while bulk hole-trap density controls the competition between 1-to-0 and 0-to-1 leakage mechanisms. Together, they explain the reduced $AC_\mathrm{min}$ for both bitflip direction.

\\ \bottomrule

\end{tabular}%
}
\end{table*}

\subsection{RowHammer}
\noindent\textbf{Single-Sided RowHammer.} There are two error mechanisms behind single-sided RowHammer: NWL-induced 1-to-0 bitflips and PWL-induced 0-to-1 bitflips. When the NWL is open, charge traps near the NWL capture electrons. These electrons are emitted when the NWL is closed, and some of these electrons migrate to the victim cell, causing a voltage drop and eventually leading to 1-to-0 bitflips. When the PWL is open, it pulls electrons away from the victim cell. When the PWL is closed, not all the electrons return to the victim cell, causing a voltage increase and eventually leading to 0-to-1 bitflips.

\noindent\textbf{Double-Sided RowHammer.} When the NWL and PWL are opened and closed in an alternating manner (i.e., double-sided RowHammer), both the 0-to-1 and 1-to-0 leakage mechanisms are enhanced (i.e., leading to much fewer aggressor row activations compared to single-sided RowHammer). When one aggressor wordline is closed, the other aggressor WL is open, pulling more electrons towards (NWL close and PWL open) or away from (PWL close and NWL open) the victim cell. Double-sided RowHammer can induce both 1-to-0 and 0-to-1 bitflips depending on the location of the charge trap. If the charge trap is near the NWL, 1-to-0 bitflips are enhanced while 0-to-1 bitflips are suppressed (vice versa for PWL). We first observe 0-to-1 bitflips (i.e., with fewer aggressor row activations than 1-to-0 bitflips) because 1) the electron migration path for the PWL-induced leakage is shorter than that for the NWL-induced leakage, and 2) capacitive coupling between the aggressor and victim WLs enhances the 0-to-1 leakage. As the aggressor row activation counts keep increasing, we eventually observe more 1-to-0 bitflips than 0-to-1 bitflips because the current DRAM device structure and manufacturing process tend to introduce more charge traps near the NWL than the PWL, which leads to more DRAM cells vulnerable to NWL-induced 1-to-0 bitflips than PWL-induced 0-to-1 bitflips.

\noindent\textbf{RowPress.} For both NWL and PWL, when they are kept open for a long period of time (i.e., RowPress), they accumulate leakage that has the opposite direction compared to RowHammer due to their strong electric field. For NWL, it enhances the off-state 0-to-1 leakage of the victim cell by weakening the gate control of the victim cell's access transistor. For PWL, it draws electrons from the substrate towards the victim cell, causing a 1-to-0 voltage shift. Compared to the RowHammer mechanisms, the RowPress mechanisms are initially weaker but their leakage accumulates over time. When the aggressor row is kept open for a sufficiently long period of time, their leakage can eventually surpass the RowHammer mechanisms, leading to opposite bitflip directions compared to RowHammer. The RowHammer leakage mechanisms are not sensitive to the increase in aggressor row on time because the charge traps saturate quickly when the aggressor row is opened. We do not observe 0-to-1 RowPress bitflips in real DRAM chip characterization because the leakage mechanism is not strong enough to overcome the data retention leakage of the victim cell in normal DRAM operating conditions. We also find that the density of hole traps in the bulk region of the device ($D_\mathrm{HT,bulk}$) is a key modeling parameter that significantly affects whether TCAD simulation results can match real DRAM chip characterization for RowPress.

\section{Implications and Future Works}
\noindent\textbf{Understanding DRAM Read Disturbance.} To our knowledge, our work is the first to comprehensively and systematically bridge the gap between experimental characterization and device-level modeling of DRAM read disturbance (i.e., RowHammer and RowPress) bitflips. Our work provides an updated and comprehensive understanding of the device-level mechanisms to explain the key RowHammer and RowPress bitflip characteristics. Our work serves as a strong foundation for follow-on works to further understand other DRAM read disturbance phenomena, including 1) variants and combinations of RowHammer and RowPress access patterns~\cite{kogler2022half, lang2023blaster, Yuksel2025PuDHammer, luo2024experimental, wang2026scaledisturb}, 2) temporal variations of read disturbance threshold~\cite{olgun2025variable}, and 3) other novel and emerging read disturbance mechanisms like ColumnDisturb~\cite{Yuksel2025ColumnDisturb}.

\noindent\textbf{System-Level Modeling of DRAM Disturbance.} Our findings highlight the complexity of the underlying error mechanisms of RowHammer and RowPress bitflips (e.g., how different access patterns involve many different and sometimes opposing and competing error mechanisms). The findings provide direct and principled guidance to build more accurate system-level DRAM read disturbance error modeling and injection frameworks beyond simple statistical bitflip models that cannot fully capture the complex characteristics of RowHammer and RowPress bitflips (e.g.,~\cite{goswami2026hammersimsystemleveltoolmodel}). We leave the development of a more accurate system-level DRAM disturbance model to future works.

\noindent\textbf{Real DRAM Characterization.} Our results also show that experimental DRAM characterization remains critical to rigorously and fully understand the real DRAM read disturbance behavior. The unveiled device-level mechanisms also provide guidelines and insights on designing more effective, rigorous, but still accurate characterization and testing methodologies, which will benefit the design, configuration, and evaluation of DRAM read disturbance mitigation techniques as well as production yield.

\section{Related Works}
To our knowledge, this is the first work to comprehensively and systematically bridge the gap between experimental characterization and device-level modeling of DRAM read disturbance (i.e., RowHammer and RowPress) bitflips.

\noindent\textbf{Experimental Characterization of DRAM Read Disturbance.} A wide range of prior works perform experimental characterization of RowHammer~\cite{kim2014flipping, kim2020revisiting, orosa2021deeper, yaglikci2022understanding, yaglikci2024svard, olgun2024read, olgun2025variable, tugrul2025understanding}, RowPress~\cite{luo2023rowpress, luo2024experimental, wang2026scaledisturb, olgun2024read, olgun2025variable}, and other forms of DRAM read disturbance~\cite{Yuksel2025PuDHammer, Yuksel2025ColumnDisturb, luo2026dejavu} using real commodity DRAM chips. However, they do not directly perform device-level modeling and simulation to understand the underlying error mechanisms of their observations. DRAMScope~\cite{nam2024dramscope} reverse engineers DRAM internal layout using DRAM read disturbance observations and insights from the $6F^2$ cell layout, but they do not identify the gaps and inconsistencies between real-chip characterization results and device-level modeling. A prior work~\cite{luo2025revisiting} identifies some gaps and inconsistencies between real-chip characterization results and device-level modeling for RowHammer and RowPress, but they do not 1) perform targeted characterizations to narrow down and pinpoint the gaps and inconsistencies to the DRAM cell array layout and specific device-level mechanisms, and 2) perform detailed TCAD device-level modeling to bridge the gap and inconsistencies.

\noindent\textbf{Device-Level Modeling of DRAM Read Disturbance.} Prior device-level works~\cite{park2014active, park2016experiments, yang2016suppression, ryu2017overcoming, yang2019trap, gautam2019row, Gautam2020MitigatingPassing, walker2021ondramrowhammer, han2021surround, zhou2023double, Zhou2024Unveiling, Zhou2024Understanding, Jie2024Understanding} perform failure analysis and TCAD simulations to identify the underlying error mechanisms of RowHammer and RowPress. However, as we show in Section~\ref{sec:exp_char} as well as in a prior work~\cite{luo2025revisiting}, they do not fully explain \emph{all} the real-chip characterization observations for RowHammer and RowPress. Our work identifies 1) the missing parts in TCAD modeling and simulation, 2) key modeling and simulation parameters that significantly affects whether simulation results match real-chip characterization results, and 3) presents updated device-level mechanisms to explain all the major bitflip characteristics of RowHammer and RowPress.

\noindent\textbf{Reverse-Engineering of DRAM Internals.} Prior works reverse engineer 1) DRAM internal organization~\cite{khan2016parbor, nam2024dramscope, marazzi2024hifidram}, 2) proprietary in-DRAM read disturbance mitigation mechanisms~\cite{frigo2020trrespass, hassan2021utrr, olgun2024read, jattke2025mcsee}, and 3) on-die ECC mechanisms~\cite{patel2019understanding, patel2020beer, patel2021harp, cojocar2019eccploit} that DRAM manufacturers do not publicly disclose. However, these works do not uncover the device-level error mechanisms of DRAM read disturbance bitflips. Our findings can inspire future DRAM reverse-engineering methodologies to leverage read disturbance bitflips to perform more accurate and fine-grained reverse engineering of DRAM internals.

\section{Conclusion}
In this paper, we comprehensively and systematically bridge the gap between experimental characterization and device-level modeling of the two most fundamental DRAM read disturbance phenomena, RowHammer and RowPress. We provide an updated and comprehensive understanding of the device-level mechanisms to explain the key RowHammer and RowPress bitflip characteristics. Our work serves as the foundation for future works to understand, characterize, and mitigate DRAM read disturbance in a principled and efficient manner.

\section*{\texorpdfstring{Acknowledgments}{Acknowledgments}}
We thank the SAFARI Research Group members for their constructive feedback and for providing a stimulating intellectual{, scholarly,} and scientific environment. We acknowledge the generous gift funding provided by our industrial partners (especially Google, Huawei, Intel, Microsoft), which has been instrumental in enabling the research we have been conducting on read disturbance in DRAM in particular and memory systems in general~\cite{kim2014flipping, mutlu2017rowhammer,mutlu2019processing,mutlu2019rowhammer,mutlu2022modern,mutlu2023fundamentally,mutlu2023retrospective,mutlu2014research,mutlu2023retrospective-raidr,mutlu2023retrospectiveexperimentalstudydata, mutlu2025memory,mutlu2013memory, kakolyris2026columnkeeper, oliveira2021damov, gomezluna2022benchmarking, ghose2019processing, mutlu2024memory, mutlu2015main, oliveira2022accelerating, singh2021fpga, cai2017flashtbd, cai2017errors, mutlu2020intelligentdate, mutlu2023tesseractretrospective, mutlu2023selfretrospective, yuksel2026memory, olgun2026drambender, bostanci2026extended}. This work was in part supported by a Google Security and Privacy Research Award and the Microsoft Swiss Joint Research Center.

\bibliographystyle{IEEEtran}
\bibliography{refs}

\begin{thebibliography}{100}
\providecommand{\url}[1]{#1}
\csname url@samestyle\endcsname
\providecommand{\newblock}{\relax}
\providecommand{\bibinfo}[2]{#2}
\providecommand{\BIBentrySTDinterwordspacing}{\spaceskip=0pt\relax}
\providecommand{\BIBentryALTinterwordstretchfactor}{4}
\providecommand{\BIBentryALTinterwordspacing}{\spaceskip=\fontdimen2\font plus
\BIBentryALTinterwordstretchfactor\fontdimen3\font minus \fontdimen4\font\relax}
\providecommand{\BIBforeignlanguage}[2]{{%
\expandafter\ifx\csname l@#1\endcsname\relax
\typeout{** WARNING: IEEEtran.bst: No hyphenation pattern has been}%
\typeout{** loaded for the language `#1'. Using the pattern for}%
\typeout{** the default language instead.}%
\else
\language=\csname l@#1\endcsname
\fi
#2}}
\providecommand{\BIBdecl}{\relax}
\BIBdecl

\bibitem{kim2014flipping}
Y.~{Kim} \emph{et~al.}, ``{Flipping Bits in Memory Without Accessing Them: An Experimental Study of DRAM Disturbance Errors},'' in \emph{ISCA}, 2014.

\bibitem{kim2020revisiting}
J.~S. Kim \emph{et~al.}, ``{Revisiting RowHammer: An Experimental Analysis of Modern DRAM Devices and Mitigation Techniques},'' in \emph{ISCA}, 2020.

\bibitem{luo2023rowpress}
H.~Luo \emph{et~al.}, ``{RowPress: Amplifying Read Disturbance in Modern DRAM Chips},'' in \emph{ISCA}, 2023.

\bibitem{luo2024experimental}
H.~Luo \emph{et~al.}, ``{An Experimental Characterization of Combined RowHammer and RowPress Read Disturbance in Modern DRAM Chips},'' in \emph{DSN Disrupt}, 2024.

\bibitem{gruss2016rowhammer}
D.~Gruss \emph{et~al.}, ``{Rowhammer.js: A Remote Software-Induced Fault Attack in JavaScript},'' in \emph{DIMVA}, 2016.

\bibitem{fournaris2017exploiting}
A.~P. Fournaris \emph{et~al.}, ``{Exploiting Hardware Vulnerabilities to Attack Embedded System Devices: A Survey of Potent Microarchitectural Attacks},'' \emph{Electronics}, 2017.

\bibitem{poddebniak2018attacking}
D.~Poddebniak \emph{et~al.}, ``{Attacking Deterministic Signature Schemes using Fault Attacks},'' in \emph{EuroS\&P}, 2018.

\bibitem{tatar2018throwhammer}
A.~Tatar \emph{et~al.}, ``{Throwhammer: {Rowhammer} {Attacks} Over the {Network} and {Defenses}},'' in \emph{{USENIX} {ATC}}, 2018.

\bibitem{carre2018openssl}
S.~Carre \emph{et~al.}, ``{OpenSSL Bellcore's Protection Helps Fault Attack},'' in \emph{DSD}, 2018.

\bibitem{barenghi2018software}
A.~Barenghi \emph{et~al.}, ``{Software-Only Reverse Engineering of Physical DRAM Mappings for Rowhammer Attacks},'' in \emph{IVSW}, 2018.

\bibitem{zhang2018triggering}
Z.~Zhang \emph{et~al.}, ``{Triggering Rowhammer Hardware Faults on ARM: A Revisit},'' in \emph{ASHES}, 2018.

\bibitem{bhattacharya2018advanced}
S.~Bhattacharya and D.~Mukhopadhyay, ``{Advanced Fault Attacks in Software: Exploiting the Rowhammer Bug},'' in \emph{Fault Tolerant Architectures for Cryptography and Hardware Security}, 2018.

\bibitem{google-project-zero}
M.~Seaborn and T.~Dullien, ``{Exploiting the DRAM Rowhammer Bug to Gain Kernel Privileges},'' \url{http://googleprojectzero.blogspot.com.tr/2015/03/exploiting-dram-rowhammer-bug-to-gain.html}, 2015.

\bibitem{rowhammergithub}
{SAFARI Research Group}, ``{RowHammer --- GitHub Repository},'' \url{https://github.com/CMU-SAFARI/rowhammer}, 2014.

\bibitem{seaborn2015exploiting}
M.~Seaborn and T.~Dullien, ``{Exploiting the DRAM Rowhammer Bug to Gain Kernel Privileges},'' \emph{Black Hat}, 2015.

\bibitem{van2016drammer}
V.~van~der Veen \emph{et~al.}, ``{Drammer: Deterministic Rowhammer Attacks on Mobile Platforms},'' in \emph{CCS}, 2016.

\bibitem{razavi2016flip}
K.~Razavi \emph{et~al.}, ``{Flip Feng Shui: Hammering a Needle in the Software Stack},'' in \emph{USENIX Security}, 2016.

\bibitem{pessl2016drama}
P.~Pessl \emph{et~al.}, ``{DRAMA: Exploiting DRAM Addressing for Cross-CPU Attacks},'' in \emph{USENIX Security}, 2016.

\bibitem{xiao2016one}
Y.~Xiao \emph{et~al.}, ``{One Bit Flips, One Cloud Flops: Cross-VM Row Hammer Attacks and Privilege Escalation},'' in \emph{USENIX Security}, 2016.

\bibitem{bosman2016dedup}
E.~Bosman \emph{et~al.}, ``{Dedup Est Machina: Memory Deduplication as An Advanced Exploitation Vector},'' in \emph{S\&P}, 2016.

\bibitem{bhattacharya2016curious}
S.~Bhattacharya and D.~Mukhopadhyay, ``{Curious Case of Rowhammer: Flipping Secret Exponent Bits Using Timing Analysis},'' in \emph{CHES}, 2016.

\bibitem{burleson2016invited}
W.~Burleson \emph{et~al.}, ``{Invited: Who is the Major Threat to Tomorrow's Security? You, the Hardware Designer},'' in \emph{DAC}, 2016.

\bibitem{qiao2016new}
R.~Qiao and M.~Seaborn, ``{A New Approach for RowHammer Attacks},'' in \emph{HOST}, 2016.

\bibitem{brasser2017can}
F.~Brasser \emph{et~al.}, ``{Can't Touch This: Software-Only Mitigation Against Rowhammer Attacks Targeting Kernel Memory},'' in \emph{USENIX Security}, 2017.

\bibitem{jang2017sgx}
Y.~Jang \emph{et~al.}, ``{SGX-Bomb: Locking Down the Processor via Rowhammer Attack},'' in \emph{SOSP}, 2017.

\bibitem{aga2017good}
M.~T. Aga \emph{et~al.}, ``{When Good Protections Go Bad: Exploiting Anti-DoS Measures to Accelerate Rowhammer Attacks},'' in \emph{HOST}, 2017.

\bibitem{mutlu2017rowhammer}
O.~Mutlu, ``{The RowHammer Problem and Other Issues We May Face as Memory Becomes Denser},'' in \emph{DATE}, 2017.

\bibitem{tatar2018defeating}
A.~Tatar \emph{et~al.}, ``{Defeating Software Mitigations Against Rowhammer: A Surgical Precision Hammer},'' in \emph{RAID}, 2018.

\bibitem{gruss2018another}
D.~Gruss \emph{et~al.}, ``{Another Flip in the Wall of Rowhammer Defenses},'' in \emph{S\&P}, 2018.

\bibitem{lipp2018nethammer}
M.~Lipp \emph{et~al.}, ``{Nethammer: Inducing Rowhammer Faults through Network Requests},'' in \emph{EuroS\&PW}, 2020.

\bibitem{van2018guardion}
V.~van~der Veen \emph{et~al.}, ``{GuardION: Practical Mitigation of DMA-Based Rowhammer Attacks on ARM},'' in \emph{{DIMVA}}, 2018.

\bibitem{frigo2018grand}
P.~Frigo \emph{et~al.}, ``{Grand Pwning Unit: Accelerating Microarchitectural Attacks with the GPU},'' in \emph{S\&P}, 2018.

\bibitem{cojocar2019eccploit}
L.~Cojocar \emph{et~al.}, ``{Exploiting Correcting Codes: On the Effectiveness of ECC Memory Against Rowhammer Attacks},'' in \emph{S\&P}, 2019.

\bibitem{ji2019pinpoint}
S.~Ji \emph{et~al.}, ``{Pinpoint Rowhammer: Suppressing Unwanted Bit Flips on Rowhammer Attacks},'' in \emph{ASIACCS}, 2019.

\bibitem{mutlu2019rowhammer}
O.~Mutlu and J.~S. Kim, ``{RowHammer: A Retrospective},'' \emph{IEEE TCAD}, 2020.

\bibitem{hong2019terminal}
S.~Hong \emph{et~al.}, ``{Terminal Brain Damage: Exposing the Graceless Degradation in Deep Neural Networks Under Hardware Fault Attacks},'' in \emph{USENIX Security}, 2019.

\bibitem{kwong2020rambleed}
A.~Kwong \emph{et~al.}, ``{RAMBleed: Reading Bits in Memory Without Accessing Them},'' in \emph{S\&P}, 2020.

\bibitem{frigo2020trrespass}
P.~Frigo \emph{et~al.}, ``{TRRespass: Exploiting the Many Sides of Target Row Refresh},'' in \emph{{S\&P}}, 2020.

\bibitem{cojocar2020rowhammer}
L.~Cojocar \emph{et~al.}, ``{Are We Susceptible to Rowhammer? An End-to-End Methodology for Cloud Providers},'' in \emph{S\&P}, 2020.

\bibitem{weissman2020jackhammer}
Z.~Weissman \emph{et~al.}, ``{JackHammer: Efficient Rowhammer on Heterogeneous FPGA--CPU Platforms},'' \emph{IACR Transactions on Cryptographic Hardware and Embedded Systems}, vol. 2020, 2020.

\bibitem{zhang2020pthammer}
Z.~Zhang \emph{et~al.}, ``{PThammer: Cross-User-Kernel-Boundary Rowhammer through Implicit Accesses},'' in \emph{MICRO}, 2020.

\bibitem{yao2020deephammer}
F.~Yao \emph{et~al.}, ``{DeepHammer: Depleting the Intelligence of Deep Neural Networks Through Targeted Chain of Bit Flips},'' in \emph{USENIX Security}, 2020.

\bibitem{deridder2021smash}
F.~de~Ridder \emph{et~al.}, ``{SMASH}: {Synchronized} {Many-Sided} {Rowhammer} {Attacks} from {JavaScript},'' in \emph{{USENIX Security}}, 2021.

\bibitem{hassan2021utrr}
H.~Hassan \emph{et~al.}, ``{Uncovering in-DRAM RowHammer Protection Mechanisms: A New Methodology, Custom RowHammer Patterns, and Implications},'' in \emph{MICRO}, 2021.

\bibitem{jattke2022blacksmith}
P.~Jattke \emph{et~al.}, ``{Blacksmith: Scalable Rowhammering in the Frequency Domain},'' in \emph{S\&P}, 2022.

\bibitem{tol2022toward}
M.~C. Tol \emph{et~al.}, ``{Don't Knock! Rowhammer at the Backdoor of DNN Models},'' in \emph{DSN}, 2023.

\bibitem{kogler2022half}
A.~Kogler \emph{et~al.}, ``{Half-Double: Hammering From the Next Row Over},'' in \emph{USENIX Security}, 2022.

\bibitem{orosa2022spyhammer}
L.~Orosa \emph{et~al.}, ``{SpyHammer: Using RowHammer to Remotely Spy on Temperature},'' {arXiv:2210.04084}, 2022.

\bibitem{zhang2022implicit}
Z.~Zhang \emph{et~al.}, ``{Implicit Hammer: Cross-Privilege-Boundary Rowhammer through Implicit Accesses},'' \emph{IEEE TDSC}, 2023.

\bibitem{liu2022generating}
L.~Liu \emph{et~al.}, ``{Generating Robust DNN with Resistance to Bit-Flip based Adversarial Weight Attack},'' \emph{IEEE TC}, 2023.

\bibitem{cohen2022hammerscope}
Y.~Cohen \emph{et~al.}, ``{HammerScope: Observing DRAM Power Consumption Using Rowhammer},'' in \emph{CCS}, 2022.

\bibitem{zheng2022trojvit}
M.~Zheng \emph{et~al.}, ``{TrojViT: Trojan Insertion in Vision Transformers},'' in \emph{CVPR}, 2023.

\bibitem{fahr2022frodo}
M.~Fahr, Jr. \emph{et~al.}, ``{When Frodo Flips: End-to-End Key Recovery on FrodoKEM via Rowhammer},'' in \emph{CCS}, 2022.

\bibitem{tobah2022spechammer}
Y.~Tobah \emph{et~al.}, ``{SpecHammer: Combining Spectre and Rowhammer for New Speculative Attacks},'' in \emph{S\&P}, 2022.

\bibitem{rakin2022deepsteal}
A.~S. Rakin \emph{et~al.}, ``{DeepSteal: Advanced Model Extractions Leveraging Efficient Weight Stealing in Memories},'' in \emph{S\&P}, 2022.

\bibitem{kang2024sledgehammer}
I.~Kang \emph{et~al.}, ``{SledgeHammer: Amplifying Rowhammer via Bank-level Parallelism},'' in \emph{USENIX Security}, 2024.

\bibitem{jattke2024zenhammer}
P.~Jattke \emph{et~al.}, ``{ZenHammer: Rowhammer Attacks on AMD Zen-based Platforms},'' in \emph{USENIX Security}, 2024.

\bibitem{bolcskei2025rubicon}
M.~B{\"o}lcskei \emph{et~al.}, ``{Rubicon: Precise Microarchitectural Attacks with Page-Granular Massaging},'' in \emph{EuroS\&P}, 2025.

\bibitem{deridder2025posthammer}
F.~de~Ridder \emph{et~al.}, ``{Posthammer: Pervasive Browser-based Rowhammer Attacks with Postponed Refresh Commands},'' in \emph{USENIX Security}, 2025.

\bibitem{jattke2025mcsee}
P.~Jattke \emph{et~al.}, ``{McSee: Evaluating Advanced Rowhammer Attacks and Defenses via Automated DRAM Traffic Analysis},'' in \emph{USENIX Security}, 2025.

\bibitem{meyer2026phoenix}
D.~Meyer \emph{et~al.}, ``{Phoenix: Rowhammer Attacks on DDR5 with Self-Correcting Synchronization},'' in \emph{S\&P}, 2026.

\bibitem{lin2025gpuhammer}
C.~S. Lin \emph{et~al.}, ``{GPUHammer: Rowhammer Attacks on GPU Memories are Practical},'' in \emph{USENIX Security}, 2025.

\bibitem{lin2026gpubreach}
C.~S. Lin \emph{et~al.}, ``{GPUBreach: Privilege Escalation Attacks on GPUs using Rowhammer},'' in \emph{S\&P}, 2026.

\bibitem{hu2026gddrhammer}
Y.~Hu \emph{et~al.}, ``{GDDRHammer: Greatly Disturbing DRAM Rows — Cross-Component Rowhammer Attacks from Modern GPUs},'' in \emph{S\&P}, 2026.

\bibitem{wan2026geforge}
J.~Wan \emph{et~al.}, ``{GeForge: Hammering GDDR Memory to Forge GPU Page Tables for Fun and Profit},'' in \emph{S\&P}, 2026.

\bibitem{shukla2026prowhammer}
M.~Shukla \emph{et~al.}, ``{PRowhammer: Propagating Bit-flips from CPU to GPU},'' in \emph{ISCA}, 2026.

\bibitem{chen2025rhammer}
W.~Chen \emph{et~al.}, ``{$\rho$Hammer: Reviving RowHammer Attacks on New Architectures via Prefetching},'' in \emph{MICRO}, 2025.

\bibitem{li2025oneflip}
X.~Li \emph{et~al.}, ``{Rowhammer-Based Trojan Injection: One Bit Flip Is Sufficient for Backdooring DNNs},'' in \emph{USENIX Security}, 2025.

\bibitem{orosa2021deeper}
L.~Orosa \emph{et~al.}, ``{A Deeper Look into RowHammer's Sensitivities: Experimental Analysis of Real DRAM Chips and Implications on Future Attacks and Defenses},'' in \emph{MICRO}, 2021.

\bibitem{yaglikci2022understanding}
A.~G. Ya{\u{g}}l{\i}k{c}{\i} \emph{et~al.}, ``{Understanding RowHammer Under Reduced Wordline Voltage: An Experimental Study Using Real DRAM Devices},'' in \emph{DSN}, 2022.

\bibitem{yaglikci2024svard}
A.~G. Ya{\u{g}}l{\i}k{\c{c}}{\i} \emph{et~al.}, ``{Spatial Variation-Aware Read Disturbance Defenses: Experimental Analysis of Real DRAM Chips and Implications on Future Solutions},'' in \emph{HPCA}, 2024.

\bibitem{nam2024dramscope}
H.~Nam \emph{et~al.}, ``{DRAMScope: Uncovering DRAM Microarchitecture and Characteristics by Issuing Memory Commands},'' in \emph{ISCA}, 2024.

\bibitem{luo2025revisiting}
H.~Luo \emph{et~al.}, ``{Revisiting DRAM Read Disturbance: Identifying Inconsistencies Between Experimental Characterization and Device-Level Studies},'' in \emph{VTS}, 2025.

\bibitem{ryu2017overcoming}
S.-W. Ryu \emph{et~al.}, ``{Overcoming the Reliability Limitation in the Ultimately Scaled DRAM using Silicon Migration Technique by Hydrogen Annealing},'' in \emph{IEDM}, 2017.

\bibitem{yang2019trap}
T.~Yang and X.-W. Lin, ``{Trap-Assisted DRAM Row Hammer Effect},'' \emph{IEEE EDL}, 2019.

\bibitem{walker2021ondramrowhammer}
A.~J. Walker \emph{et~al.}, ``{On DRAM RowHammer and the Physics of Insecurity},'' \emph{IEEE TED}, 2021.

\bibitem{Gautam2020MitigatingPassing}
S.~K. Gautam \emph{et~al.}, ``{Mitigating the Passing Word Line Induced Soft Errors in Saddle Fin DRAM},'' \emph{IEEE TED}, 2020.

\bibitem{zhou2023double}
L.~Zhou \emph{et~al.}, ``{Double-sided Row Hammer Effect in Sub-20 nm DRAM: Physical Mechanism, Key Features and Mitigation},'' in \emph{IRPS}, 2023.

\bibitem{Zhou2024Unveiling}
L.~Zhou \emph{et~al.}, ``{Unveiling RowPress in Sub-20 nm DRAM Through Comparative Analysis With Row Hammer: From Leakage Mechanisms to Key Features},'' in \emph{IEEE Transactions on Electron Devices}, 2024.

\bibitem{Zhou2024Understanding}
L.~Zhou \emph{et~al.}, ``{Understanding the Physical Mechanism of RowPress at the Device-Level in Sub-20 nm DRAM},'' in \emph{IRPS}, 2024.

\bibitem{olgun2023dram}
A.~Olgun \emph{et~al.}, ``{DRAM Bender: An Extensible and Versatile FPGA-based Infrastructure to Easily Test State-of-the-art DRAM Chips},'' \emph{IEEE TCAD}, 2023.

\bibitem{TCAD2018manual}
{Synopsys}, ``{TCAD Sentaurus Manual},'' Mountain View, CA, USA, 2018.

\bibitem{Park2006SFinRCAT}
S.-W. Park \emph{et~al.}, ``{Highly Scalable Saddle-Fin (S-Fin) Transistor for Sub-50nm DRAM Technology},'' in \emph{Symposium on VLSI Technology}, 2006.

\bibitem{Chae2024SingleMetalBCAT}
K.~Chae \emph{et~al.}, ``{Single Metal BCAT Breakthrough to Open a New Era of 12 nm DRAM and Beyond},'' in \emph{Symposium on VLSI Technology and Circuits}, 2024.

\bibitem{Yang2013SuperiorImprovements}
C.-M. Yang \emph{et~al.}, ``{Superior Improvements in GIDL and Retention by Fluorine Implantation in Saddle-Fin Array Devices for Sub-40-nm DRAM Technology},'' \emph{IEEE EDL}, 2013.

\bibitem{Jie2024Understanding}
J.~Li \emph{et~al.}, ``{Understanding the Competitive Interaction in Leakage Mechanisms for Effective Row Hammer Mitigation in Sub-20 nm DRAM},'' \emph{IEEE EDL}, 2024.

\bibitem{Jang2025Gateoxidetechnology}
D.~Jang \emph{et~al.}, ``{Gate oxide technology relieving word-line break in 10 nm-class DRAMs},'' \emph{Japanese Journal of Applied Physics}, 2025.

\bibitem{Jeon2017InvestigationOnTheLocal}
S.~Jeon \emph{et~al.}, ``{Investigation on the Local Variation in BCAT Process for DRAM Technology},'' in \emph{IRPS}, 2017.

\bibitem{lang2023blaster}
Z.~Lang \emph{et~al.}, ``{{BLASTER}: Characterizing the Blast Radius of Rowhammer},'' in \emph{DRAMSec}, 2023.

\bibitem{Yuksel2025PuDHammer}
I.~E. Yuksel \emph{et~al.}, ``{PuDHammer: Experimental Analysis of Read Disturbance Effects of Processing-using-DRAM in Real DRAM Chips},'' in \emph{ISCA}, 2025.

\bibitem{wang2026scaledisturb}
J.~Wang \emph{et~al.}, ``{ScaleDisturb: Exploiting Temporal Asymmetry to Amplify Read Disturbance in Modern DRAM Chips},'' in \emph{DSN}, 2026.

\bibitem{olgun2025variable}
A.~Olgun \emph{et~al.}, ``{Variable Read Disturbance: An Experimental Analysis of Temporal Variation in DRAM Read Disturbance},'' in \emph{HPCA}, 2025.

\bibitem{Yuksel2025ColumnDisturb}
I.~E. Yuksel \emph{et~al.}, ``{ColumnDisturb: Understanding Column-based Read Disturbance in Real DRAM Chips and Implications for Future Systems},'' in \emph{MICRO}, 2025.

\bibitem{goswami2026hammersimsystemleveltoolmodel}
K.~Goswami \emph{et~al.}, ``{HammerSim: A System-Level Tool to Model RowHammer},'' {arXiv:2605.27803}, 2026.

\bibitem{olgun2024read}
A.~Olgun \emph{et~al.}, ``{Read Disturbance in High Bandwidth Memory: A Detailed Experimental Study on HBM2 DRAM Chips},'' in \emph{DSN}, 2024.

\bibitem{tugrul2025understanding}
Y.~C. Tugrul \emph{et~al.}, ``{Understanding RowHammer Under Reduced Refresh Latency: Experimental Analysis of Real DRAM Chips and Implications on Future Solutions},'' in \emph{HPCA}, 2025.

\bibitem{luo2026dejavu}
H.~Luo \emph{et~al.}, ``{DejaVu: Why You Should Write to Your DRAM Rows Twice, Carefully},'' in \emph{ISCA}, 2026.

\bibitem{park2014active}
K.~Park \emph{et~al.}, ``{Active-Precharge Hammering on a Row-Induced Failure in DDR3 SDRAMs Under 3x nm Technology},'' in \emph{IIRW}, 2014.

\bibitem{park2016experiments}
K.~Park \emph{et~al.}, ``{Experiments and Root Cause Analysis for Active-Precharge Hammering Fault in DDR3 SDRAM under 3xnm Technology},'' \emph{Microelectronics Reliability}, 2016.

\bibitem{yang2016suppression}
C.~Yang \emph{et~al.}, ``{Suppression of RowHammer Effect by Doping Profile Modification in Saddle-Fin Array Devices for Sub-30-nm DRAM Technology},'' \emph{TDMR}, 2016.

\bibitem{gautam2019row}
S.~Gautam \emph{et~al.}, ``{Row Hammering Mitigation Using Metal Nanowire in Saddle Fin DRAM},'' \emph{IEEE TED}, 2019.

\bibitem{han2021surround}
J.-W. Han \emph{et~al.}, ``{Surround Gate Transistor With Epitaxially Grown Si Pillar and Simulation Study on Soft Error and Rowhammer Tolerance for DRAM},'' \emph{IEEE TED}, 2021.

\bibitem{khan2016parbor}
S.~Khan \emph{et~al.}, ``{PARBOR: An Efficient System-Level Technique to Detect Data-Dependent Failures in DRAM},'' in \emph{DSN}, 2016.

\bibitem{marazzi2024hifidram}
M.~Marazzi \emph{et~al.}, ``{HiFi-DRAM: Enabling High-fidelity DRAM Research by Uncovering Sense Amplifiers with IC Imaging},'' in \emph{ISCA}, 2024.

\bibitem{patel2019understanding}
M.~Patel \emph{et~al.}, ``{Understanding and Modeling On-Die Error Correction in Modern DRAM: An Experimental Study Using Real Devices},'' in \emph{DSN}, 2019.

\bibitem{patel2020beer}
M.~Patel \emph{et~al.}, ``{Bit-Exact ECC Recovery (BEER): Determining DRAM On-Die ECC Functions by Exploiting DRAM Data Retention Characteristics},'' in \emph{{MICRO}}, 2020.

\bibitem{patel2021harp}
M.~Patel \emph{et~al.}, ``{HARP: Practically and Effectively Identifying Uncorrectable Errors in Memory Chips That Use On-Die Error-Correcting Codes},'' in \emph{MICRO}, 2021.

\bibitem{mutlu2019processing}
O.~Mutlu \emph{et~al.}, ``{Processing Data Where It Makes Sense: Enabling In-Memory Computation},'' in \emph{Microprocessors and Microsystems}, 2019.

\bibitem{mutlu2022modern}
O.~Mutlu \emph{et~al.}, ``{A Modern Primer on Processing in Memory},'' in \emph{Emerging computing: from devices to systems: looking beyond Moore and Von Neumann}.\hskip 1em plus 0.5em minus 0.4em\relax Springer, 2022.

\bibitem{mutlu2023fundamentally}
O.~Mutlu \emph{et~al.}, ``{Fundamentally Understanding and Solving RowHammer},'' in \emph{ASP-DAC}, 2023.

\bibitem{mutlu2023retrospective}
O.~Mutlu, ``{Retrospective: Flipping Bits in Memory without Accessing Them: An Experimental Study of DRAM Disturbance Errors},'' \emph{arXiv:2306.16093}, 2023.

\bibitem{mutlu2014research}
O.~Mutlu and L.~Subramanian, ``{Research Problems and Opportunities in Memory Systems},'' \emph{SUPERFRI}, 2014.

\bibitem{mutlu2023retrospective-raidr}
O.~Mutlu, ``{Retrospective: RAIDR: Retention-Aware Intelligent DRAM Refresh},'' \emph{arXiv:2306.16024}, 2023.

\bibitem{mutlu2023retrospectiveexperimentalstudydata}
O.~Mutlu, ``{Retrospective: An Experimental Study of Data Retention Behavior in Modern DRAM Devices: Implications for Retention Time Profiling Mechanisms},'' \emph{arXiv:2306.16037}, 2023.

\bibitem{mutlu2025memory}
O.~Mutlu \emph{et~al.}, ``{Memory-Centric Computing: Solving Computing's Memory Problem},'' in \emph{IMW}, 2025.

\bibitem{mutlu2013memory}
O.~Mutlu, ``{Memory Scaling: A Systems Architecture Perspective},'' in \emph{IMW}, 2013.

\bibitem{kakolyris2026columnkeeper}
A.~K. Kakolyris \emph{et~al.}, ``{ColumnKeeper: Efficient Solutions to the ColumnDisturb Vulnerability in DRAM-based Systems},'' in \emph{ISCA}, 2026.

\bibitem{oliveira2021damov}
G.~F. Oliveira \emph{et~al.}, ``{DAMOV: A New Methodology and Benchmark Suite for Evaluating Data Movement Bottlenecks},'' \emph{IEEE Access}, 2021.

\bibitem{gomezluna2022benchmarking}
J.~G\'{o}mez-Luna \emph{et~al.}, ``{Benchmarking a New Paradigm: Experimental Analysis and Characterization of a Real Processing-in-Memory System},'' \emph{IEEE Access}, 2022.

\bibitem{ghose2019processing}
S.~Ghose \emph{et~al.}, ``{Processing-in-Memory: A Workload-Driven Perspective},'' in \emph{IBM J. Res. Dev.}, 2019.

\bibitem{mutlu2024memory}
O.~Mutlu \emph{et~al.}, ``{Memory-Centric Computing: Recent Advances in Processing-in-DRAM (Invited)},'' in \emph{IEDM}, 2024.

\bibitem{mutlu2015main}
O.~Mutlu \emph{et~al.}, ``{The Main Memory System: Challenges and Opportunities},'' \emph{Communications of the KIISE}, 2015.

\bibitem{oliveira2022accelerating}
G.~F. Oliveira \emph{et~al.}, ``{Accelerating Neural Network Inference with Processing-in-DRAM: From the Edge to the Cloud},'' \emph{IEEE Micro}, 2022.

\bibitem{singh2021fpga}
G.~Singh \emph{et~al.}, ``{FPGA-Based Near-Memory Acceleration of Modern Data-Intensive Applications},'' \emph{IEEE Micro}, 2021.

\bibitem{cai2017flashtbd}
Y.~Cai \emph{et~al.}, ``{Error Characterization, Mitigation, and Recovery in Flash Memory Based Solid-State Drives},'' \emph{Proc. IEEE}, 2017.

\bibitem{cai2017errors}
Y.~Cai \emph{et~al.}, ``{Errors in Flash-Memory-Based Solid-State Drives: Analysis, Mitigation, and Recovery},'' \emph{arXiv preprint arXiv:1711.11427}, 2017.

\bibitem{mutlu2020intelligentdate}
O.~Mutlu, ``{Intelligent Architectures for Intelligent Computing Systems},'' in \emph{DATE}, 2021.

\bibitem{mutlu2023tesseractretrospective}
J.~Ahn \emph{et~al.}, ``{Retrospective: A Scalable Processing-in-Memory Accelerator for Parallel Graph Processing},'' Retrospective Issue for ISCA-50, 2023.

\bibitem{mutlu2023selfretrospective}
J.~F. Mart{\'i}nez \emph{et~al.}, ``{Retrospective: Self-optimizing Memory Controllers: A Reinforcement Learning Approach},'' Retrospective Issue for ISCA-50, 2023.

\bibitem{yuksel2026memory}
I.~E. Yuksel \emph{et~al.}, ``{Memory-Centric Computing: Security Benefits and Challenges of Processing-in-DRAM},'' The 7th Workshop on Memory-Centric Computing Systems colocated with ICS, 2026.

\bibitem{olgun2026drambender}
A.~Olgun \emph{et~al.}, ``{A Modern Large-Scale Memory Characterization Laboratory},'' The 3rd Tutorial on Ramulator and DRAM Bender colocated with ICS, 2026.

\bibitem{bostanci2026extended}
F.~N. Bostanc{\i} \emph{et~al.}, ``{Extended Abstract: Re-Evaluating the Real-System Modeling Accuracy of Ramulator 2.0},'' The 3rd Tutorial on Ramulator and DRAM Bender colocated with ICS, 2026.

\end{thebibliography}

\end{document}